%% file: 000-main.tex
\newif\ifnotready\notreadyfalse
\newif\ifready\readytrue
\newif\iffullpaper\fullpapertrue

\documentclass[11pt]{article}

\usepackage[utf8]{inputenc}
\usepackage[top=1in,left=1in,right=1in,bottom=1in]{geometry}
\usepackage{amsthm}
\usepackage{amsmath}
\usepackage{amssymb}
\usepackage{hyperref}
\usepackage{algorithm}
\usepackage{relsize}
\usepackage[noend]{algpseudocode}
\usepackage{algorithmicx}
\usepackage{multicol}
\usepackage{relsize}
\usepackage[most]{tcolorbox}
\usepackage{color}
\usepackage{thm-restate}
\usepackage{mathtools}
\usepackage{enumerate}
\usepackage{pgfplots}
\usepackage{subcaption}
\usepackage{graphicx}
\usepackage{booktabs}
\usepackage{makecell}
\pgfplotsset{
    compat=1.3,
    legend image code/.code={
        \draw [#1] (0cm,-0.1cm) rectangle (0.6cm,0.1cm);
    },
}
\usepackage[capitalize, nameinlink]{cleveref}

\theoremstyle{plain}

\newtheorem{theorem}{Theorem}[section]

\newtheorem{lemma}[theorem]{Lemma}
\newtheorem{corollary}[theorem]{Corollary}
\newtheorem{definition}[theorem]{Definition}

\newtheorem{notation}[theorem]{Notation}

\newtheorem{challenge}[lemma]{Challenge}

\input{macro}

\input{utils}

\crefname{theorem}{Theorem}{Theorems}
\Crefname{lemma}{Lemma}{Lemmas}
\Crefname{claim}{Claim}{Claims}
\Crefname{observation}{Observation}{Observations}
\Crefname{algorithm}{Algorithm}{Algorithms}
\Crefname{myalgctr}{Algorithm}{Algorithms}
\Crefname{challenge}{Challenge}{Challenges}

\ifnotready

\usepackage{todonotes}

\fi

\ifready

\usepackage[disable]{todonotes}
\fi
\ifnotready

\fi

\newcommand{\eqdef}{\stackrel{\text{\tiny\rm def}}{=}}

\newcommand{\floor}[1]{\lfloor{#1\rfloor}}

\newcommand{\davg}{d_{avg}}
\newcommand{\pV}{\hat{p}}
\newcommand{\triangles}[1]{\Delta(#1)}

\algrenewcommand\algorithmicindent{1em}%

\newcommand{\mM}{\mathcal{M}}

\newcommand{\cY}{\mathcal{Y}}
\newcommand{\tO}{\widetilde{O}}
\newcommand{\rb}[1]{\left( #1 \right)}
\newcommand{\ee}[1]{\mathbb{E} \left[ #1 \right]}
\newcommand{\pp}[1]{\mathbb{P} \left[ #1 \right]}
\newcommand{\var}[1]{\mathrm{Var} \left[ #1 \right]}

\newcommand{\sub}{K}
\newcommand{\countest}{B}
\newcommand{\cntSub}{H_{\#}}

\newcommand{\eat}[1]{{\color{red}{}}}

\DeclarePairedDelimiter\ceil{\lceil}{\rceil}
\newcommand{\myparagraph}[1]{\noindent {\bf #1.}}

\ifnotready
\author{Amartya Shankha Biswas}
\affiliation{%
  \institution{CSAIL, MIT}
  \city{Cambridge, MA}
  \country{USA}}
\email{asbiswas@mit.edu}

\author{Talya Eden}
\affiliation{%
  \institution{CSAIL, MIT}
  \city{Cambridge, MA}
  \country{USA}}
\email{teden@mit.edu}

\author{Quanquan C. Liu}
\affiliation{%
  \institution{CSAIL, MIT}
  \city{Cambridge, MA}
  \country{USA}}
\email{quanquan@mit.edu}

\author{Slobodan Mitrovi\'{c}}
\affiliation{%
  \institution{CSAIL, MIT}
  \city{Cambridge, MA}
  \country{USA}}
\email{slobo@mit.edu}

\author{Ronitt Rubinfeld}
\affiliation{%
  \institution{CSAIL, MIT}
  \city{Cambridge, MA}
  \country{USA}}
\email{ronitt@csail.mit.edu}
\fi
\author{Amartya Shankha Biswas\thanks{
    CSAIL, MIT, Cambridge, MA, USA,
    \texttt{\{asbiswas,slobo\}@mit.edu, ronitt@csail.mit.edu}}
    \qquad\quad
    Talya Eden\thanks{
    Boston University/CSAIL, MIT, Cambridge, MA, USA,
    \texttt{teden@mit.edu}}
     \\[0.5em]
    Quanquan C. Liu\thanks{Northwestern University, Evanston, IL, USA,\texttt{\{quanquan\}@northwestern.edu}}\qquad\quad
    Slobodan Mitrovi\'{c}\thanks{UC Davis, Davis, CA, \texttt{smitrovic@ucdavis.edu}}
    \qquad\quad
    Ronitt Rubinfeld\footnotemark[1]%
}

\begin{document}
\title{Massively Parallel Algorithms for Small Subgraph Counting}

\ifnotready
\fi

\maketitle

\begin{abstract}
Over the last two decades, frameworks for distributed-memory
parallel computation, such as
MapReduce, Hadoop, Spark and Dryad, have gained significant popularity with the
growing prevalence of large network datasets.
The Massively Parallel Computation (MPC) model is the de-facto standard for studying
graph algorithms in these frameworks theoretically.
Subgraph counting is one such fundamental problem in analyzing massive graphs, with the
main algorithmic challenges centering on designing methods which are both
scalable and accurate.

Given a graph $G=(V, E)$ with $n$ vertices, $m$ edges and $T$ triangles,
our first result is an algorithm that
outputs a $(1+\eps)$-approximation to $T$, with asymptotically \emph{optimal round and total space complexity}
provided any $S \geq \max{(\sqrt m, n^2/m)}$ space per machine and assuming $T=\Omega(\sqrt{m/n})$.
Our result gives a quadratic improvement on the bound on $T$ over previous
works. We also provide a simple extension of our result to counting \emph{any} subgraph
of $k$ size for constant $k \geq 1$.
Our second result is an $\hideo(\log \log n)$-round algorithm for exactly
counting the number of triangles, whose total space usage is
parametrized by the \emph{arboricity} $\alpha$
of the input graph. 	\eat{The space per machine is $O(n^{\delta})$ for any
constant $\delta> 0$, and the total space is $O(m\alpha)$, which matches the
\emph{time} complexity of (combinatorial) triangle counting in the sequential
model.} We extend this result to exactly counting $k$-cliques for any constant
$k$.
Finally, we prove that a recent result of Bera, Pashanasangi and Seshadhri
(ITCS 2020) for exactly counting all subgraphs of size at most $5$ can be
implemented in the MPC model in
$\tilde{O}_{\delta}(\sqrt{\log n})$ rounds, $O(n^{\delta})$ space per machine and $O(m\alpha^3)$
 total space.

 In addition to our theoretical results, we simulate our triangle counting
 algorithms
 in real-world graphs obtained from the Stanford Network Analysis Project
 (SNAP) database.
 Our results show that both our approximate and exact counting algorithms
 exhibit improvements in terms of round complexity and approximation ratio,
 respectively, compared to two previous widely used algorithms for these problems.
\end{abstract}

\input{100-introduction}
\input{120-related-work}

\input{200-preliminaries}

\section{Overview of Our Techniques}

\input{220-overview_exact}

\input{230-overview_estimator}
\input{240-overview_five_subgraphs}

\input{260-triangle_counting}

\input{450-exact-appendix}
\input{300-triangle-estimates}

\input{305-estimator_challenges}
    \input{310-fit_on_machine}
    \input{315-induced_subgraph_hashing_protocol_challenge}

    \input{660-concentration-of-counts}

    \input{320-protocol_sub_challenge}
    \input{620-bounding_light_edges}
    \input{630-bounding_heavy_edges}

    \input{640-sent_messages}
    \input{330-num_triangles_challenge}

    \input{680-high_probability_bound}
    \input{k-subgraph}

\input{400-five-subgraph}

\input{590-experiments}
\input{700-experiments-appendix}

\section{Open Questions}\label{sec:open}
There are many interesting open questions that result from our study; among these open questions
include improving the bounds presented in our algorithm: the round complexity and total
space usage in our exact algorithms and the space per machine in our approximation algorithms. In addition to
these questions, we also discussion two additional open questions with a larger research scope.

\paragraph{Small subgraph counting counting for a broader class of small subgraphs} Two recent works of~\cite{bressan2019faster,bera2021near} extend the result of~\cite{BPS20} to a broader set of small subgraphs
in the sequential model. However, their results depend crucially on a \emph{DAG tree decomposition} which is
non-trivial to implement in the MPC model. Furthermore, even given this DAG tree decomposition, their approach
requires iterating through the tree from the leaf level by level up
the tree. Such a procedure when implemented in the
MPC model requires number of rounds that is $O(depth)$ where $depth$ is the depth of the tree. The depth may not
be $\poly(\log n)$. In order to obtain efficient MPC implementation of these new algorithms, we must find
novel solutions to the above two challenges.

\paragraph{Counting in the AMPC model} A new (stronger) model of MPC, called the \emph{adaptive} MPC model,
was recently introduced by~\cite{behnezhad2021massively}. The AMPC model allows access to a \emph{shared} distributed hash table at the end of every round; additionally, the algorithms are allowed \emph{adaptive}
access to this hash table. Such a model has shown to be very practical and have led to improvements in
the number of rounds over previous MPC algorithms. Such a model seems to be quite relevant to our work
since one of the main challenges in our approximation algorithms is to find the set of edges to give to each
machine. (Such a challenge may no longer exist given a shared-memory distributed hash table.) We leave as an
interesting open question to obtain better, more round efficient approximate triangle counting algorithms in
the AMPC model.

\paragraph{Triangle Counting in $O(1)$ Rounds in Sparse Graphs}
For sparse graphs where $m = \tilde{O}(n)$,
our approximation algorithm requires $\tilde{\Omega}(n)$ space per machine which means
that (almost) the entire graph can fit on one machine. This naturally leads to an interesting
open question for whether we can obtain an approximate or exact
triangle counting algorithm in $O(1)$ rounds in sparse
graphs while using \emph{sublinear} space per machine ($n^{\delta}$ space for any constant $\delta > 0$).

\ifnotready
\section*{Acknowledgements}
S.~Mitrovi\' c was supported by the Swiss NSF grant No.~P400P2\_191122/1 and FinTech@CSAIL.
\fi

\bibliographystyle{alpha}
\bibliography{ref}

\appendix
\input{1200-appendix-preliminaries}

\end{document}

%% file: macro.tex
\newcommand{\eps}{\varepsilon}

\newcommand{\hideo}{O_{\delta}}

\newcommand{\MPC}[0]{\ensuremath{\mathsf{MPC}}}

\newcommand{\scanpram}[0]{\emph{scan} \ensuremath{\mathsf{PRAM}}\renewcommand{\scanpram}[0]{scan \ensuremath{\mathsf{PRAM}}}}
\newcommand{\arbpram}[0]{\emph{arbitrary} CRCW \ensuremath{\mathsf{PRAM}}\renewcommand{\arbpram}[0]{arbitrary \ensuremath{\mathsf{PRAM}}}}
\newcommand{\pripram}[0]{\emph{priority} CRCW \ensuremath{\mathsf{PRAM}}\renewcommand{\pripram}[0]{priority CRCW \ensuremath{\mathsf{PRAM}}}}
\newcommand{\combpram}[0]{\emph{combining} CRCW \ensuremath{\mathsf{PRAM}}\renewcommand{\combpram}[0]{combining CRCW \ensuremath{\mathsf{PRAM}}}}

\newcommand{\multipram}[0]{\emph{multiprefix} CRCW \ensuremath{\mathsf{PRAM}}\renewcommand{\multipram}[0]{multiprefix CRCW \ensuremath{\mathsf{PRAM}}}}

\algblock{ParFor}{EndParFor}
\algnewcommand\algorithmicparfor{\textbf{parfor}}
\algnewcommand\algorithmicpardo{\textbf{do}}
\algnewcommand\algorithmicendparfor{\textbf{end\ parfor}}
\algrenewtext{ParFor}[1]{\algorithmicparfor\ #1\ \algorithmicpardo}
\algrenewtext{EndParFor}{\algorithmicendparfor}

%% file: utils.tex
\DeclareMathOperator{\poly}{poly}

\definecolor{mygreen}{RGB}{20,140,80}
\definecolor{linkcolor}{RGB}{0,0,230}
\definecolor{mylightgray}{RGB}{230,230,230}
\definecolor{verylightgray}{RGB}{245,245,245}

\hypersetup{
     colorlinks=true,
     citecolor= mygreen,
     linkcolor= mygreen,
     urlcolor= mygreen
}

\newcounter{myalgctr}

\newtcolorbox{OuterBox}[1][]{%
    breakable,
    enhanced,
    frame hidden,
    interior hidden,
    left=-5pt,
    right=-5pt,
    top=-5pt,
    float=p,
    boxsep=0pt,
    arc=0pt
#1}%

\newtcolorbox{InnerBox}[1][]{%
    enforce breakable,
    enhanced,
    colback=gray,
    colframe=white,
#1}%

\newenvironment{tbox}{
\vspace{0.2cm}
\begin{tcolorbox}[width=\columnwidth,
                  enhanced,
                  boxsep=2pt,
                  left=1pt,
                  right=1pt,
                  top=4pt,
                  boxrule=1pt,
                  arc=0pt,
                  colback=white,
                  colframe=black,
	              breakable
                  ]%
}{
\end{tcolorbox}
}

\newcommand{\tboxhrule}[0]{\vspace{0.1cm} {\color{black} \hrule} \vspace{0.2cm}}

\newenvironment{titledtbox}[1]{\begin{tbox}#1 \tboxhrule}{\end{tbox}}

\newenvironment{tboxalg}[1]{\refstepcounter{myalgctr}\begin{titledtbox}{\textbf{Algorithm \themyalgctr.} #1}}{\end{titledtbox}}

%% file: 100-introduction.tex
\section{Introduction}

\sloppy
Estimating the number of small subgraphs, cliques in particular, is a fundamental problem in computer science, and has been extensively studied both theoretically and from an applied perspective.
Given its importance, the task of counting subgraphs has been explored in various computational settings, e.g., sequential~\cite{alon1997finding,V09, Bjorklund09}, distributed and parallel~\cite{suri2011counting,pagh2012colorful,kolda2014counting,park2014mapreduce,lai2015scalable},
streaming~\cite{bar2002reductions,kane2012counting,bera2017towards,mcgregor2016better},
and sublinear-time~\cite{ELRS15, Aliak,AKK,ERS20}. There are usually two perspectives from which
subgraph counting is studied: first, optimizing the running time
(especially relevant in the sequential and sublinear-time settings) and, second, optimizing the space or query
requirement (relevant in the streaming, parallel, and distributed settings).
In each of these perspectives, there are two, somewhat orthogonal, directions that one can take.
The first is \emph{exact} counting.
However, in most scenarios, algorithms that perform exact counting are prohibitive,
e.g., they require too much space or too many parallel rounds to be implementable in practice.

Hence, the second direction of obtaining an \emph{estimate/approximation}
on the number of small subgraphs is both an interesting theoretical problem and of practical importance. If $\cntSub$ is the number of subgraphs isomorphic to $H$, the main question in approximate counting is whether we can
design algorithms that, under given resource constraints, provide approximations that concentrate well. This concentration is usually parametrized by $\cntSub$ (and potentially some other parameters). In particular, most known results do not provide a strong approximation guarantee when $\cntSub$ is very small, e.g., $|\cntSub| = O(1)$. So, the main attempts in this line of work is to provide an estimation that concentrates well while imposing as small a lower bound on $\cntSub$ as possible.

Due to ever increasing sizes of data stores,
there has been an increasing interest in designing scalable algorithms.
The \emph{Massively Parallel Computation} (MPC) model is a theoretical abstraction of popular frameworks for large-scale computation such as MapReduce \cite{dean2008mapreduce}, Hadoop \cite{white2012hadoop}, Spark \cite{zaharia2010spark} and Dryad \cite{isard2007dryad}. MPC gained significant interest recently, most prominently in building algorithmic toolkits for graph processing~\cite{GSZ11, LattanziMSV11, Beame13, Andoni:2014, Beame14, hegeman2015lessons, AhnGuha15,Roughgarden16,Im17,czumaj2017round, assadi2017simple,assadi2017coresets,ghaffari2018improved,harvey2018greedy,brandt2018breaking,assadi2018massively,boroujeni2018approximating,BehnezhadDHKS19, Behnezhad0DFHKU19,BehnezhadHH19,AndoniSZ19,AssadiSW19,GLM19,GamlathKMS19,GhaffariU19, LackiMOS20,ItalianoLMP19,ChangFGUZ19,GhaffariKU19,Ghaffari0T20}. Efficiency of an algorithm in MPC is characterized by three parameters: round complexity, the space per machine in the system, and the number of machines/total memory used.
Our work aims to design efficient algorithms with respect to all three parameters and is guided by the following question:
\begin{center}
    \emph{How does one design efficient massively parallel algorithms for small subgraph counting?}
\end{center}

\subsection{The MPC Model}
In this paper, we are working in the Massively Parallel Computation (MPC) model introduced by~\cite{karloff2010MapReduce,GSZ11, Beame13}. The model
operates as follows. There exist $\mM$ machines that communicate with each other in synchronous rounds. The graph
input is initially distributed across the machines in some organized way such that machines know how to access the
relevant information via communication with other machines. During each round, the machines first perform computation
locally without communicating with other machines. The computation done locally can be unbounded (although the machines
have limited space so any reasonable program will not do an absurdly large amount of computation). At the end of the round,
the machines exchange messages to inform the computation for the next round.
The total size of all messages that can be received by a machine is upper bounded by
the size of its local memory, and each machine outputs
messages of sufficiently small size that can fit into its memory.
If $N$ is the total size of the data and each machine has $S$ words of space, we are interested in the settings when $S$ is sublinear
in $N$. We use \emph{total space} to refer to $\mM \cdot S$, which is the
total space that is available across all the machines.

\subsection{Our Contributions}

\begin{table}[htb!]
    \centering
    \footnotesize
    \begin{tabular}{| c | c | c | c | c | c |}
    \toprule
         Problem & Work & MPC Rounds & Space Per Machine & Total Space & ALB\\
         \hline
         \hline
         Exact Triangle Counting & \makecell{\cite{suri2011counting} \\ \cite{suri2011counting} \\ \cite{chu2011triangle} \\ folklore \\ \textbf{Ours}} & \makecell{$2$ \\ $1$ \\ $O(n)$ \\ $O(\log n)$ \\ \textbf{$\mathbf{O_{\delta}(\log \log n)}$}} & \makecell{$O(\sqrt{m})$ \\ $o(m)$ \\ $O(n)$ \\ $\Omega(\alpha^2)$ \\ \textbf{$\mathbf{O(n^{\delta})}$}} &
         \makecell{$O(m^{3/2})$ \\ $\omega(m)$ \\ $O(m)$ \\
         $O(m\alpha)$ \\ $\mathbf{O(m\alpha)}$} &
         \makecell{- \\ - \\ - \\ - \\ -}\\
         \hline
         \hline
         Approximate Triangle Counting & \makecell{\cite{pagh2012colorful} \\ \cite{seshadhri2013triadic} \\ \textbf{Ours}} &
         \makecell{$O(1)$ \\ $O(1)$ \\ \textbf{$\mathbf{O(1)}$}} &
         \makecell{$\Omega(m)$ \\ $O(n^{\delta})$ \\ \textbf{$\mathbf{\tO(n)}$}} &
         \makecell{$O(m)$ \\ $O(m)$ \\ \textbf{$\mathbf{\tO(m)}$}}
         & \makecell{$\Omega(d_{avg})$ \\ $\Omega\left(\sum_{v \in V} \deg(v)^2\right)$ \\
         \textbf{$\mathbf{\Omega(\sqrt{d_{avg}})}$}} \\
    \bottomrule
    \end{tabular}
    \caption{Summary of our main MPC triangle counting results compared to previous work. Our
    results are \textbf{bolded}. ``ALB'' refers to the approximation
    lower bound on the number of triangles required to
    obtain a $(1+\eps)$-approximation, with high probability. $\alpha$
    is the arboricity of the input graph and is generally small (logarithmic)
    in real-world networks. Parameter $\delta > 0$ is \emph{any} constant.}
    \label{table:results}
\end{table}

\subsubsection{Triangle Counting}
We provide a number of results for triangle counting in both the approximate and exact settings.
Let $G=(V, E)$ be a graph with $n$ vertices, $m$ edges and $T$ triangles.
First we study the question of approximately counting the number of triangles
under the restriction that the round and total space complexities are
essentially \emph{optimal}, i.e., $O(1)$ and $\tO(m)$, where $\tilde{O}$
hides $O(\poly \log n)$ factors, respectively.
Here and throughout, we use $\hideo$ and $O_{\eps}$ to hide
factors of $\delta$ and $\eps$, respectively, where we consider constant factors of
$\delta, \eps > 0$ in this paper.

Our algorithm is surprisingly simple with a
more complicated analysis, but improves on the previous best-known result by giving
a $(1+\eps)$-approximation, with high probability, while achieving a \emph{quadratic}
improvement on the number of triangles required to ensure this approximation.
The specific bounds are given in~\cref{table:results}.

\begin{restatable}{theorem}{EstimatorGrand}\label{thm:approximate-counting}
	Let $G=(V, E)$ be a graph with $n$ vertices, $m$ edges, and let $T$ be the number of triangles in $G$.
	Assuming
	\begin{multicols}{2}
		\begin{enumerate}[(i)]
			\item $T = \widetilde\Omega\left( \sqrt{\frac{m}{S}}\right)$,
			\item $S = \widetilde{\Omega}\left( \max\left\{\frac{\sqrt{m}}{\eps}, \frac{n^2}{m} \right\} \right)$,
		\end{enumerate}
	\end{multicols}
	there exists an $\MPC$ algorithm, using $\mM$ machines, each with local space $S$, and total space $\mM S = \tilde O_{\eps}(m)$,
	that outputs a $(1\pm \eps)$-approximation of $T$, with high probability, in $O(1)$ rounds.
\end{restatable}

For $S = \Theta(n\log n)$ (specifically, $S > 100n\log n$) in \cref{thm:approximate-counting}, we derive the following corollary.
\begin{corollary}
	Let $G$ be a graph and $T$ be the number of triangles it contains. If $T \ge \sqrt{d_{avg}}$,
    then there exists an MPC algorithm that in $O(1)$ rounds with high
    probability outputs a $(1 + \eps)$-approximation of $T$. This algorithm uses
    a total space of $\tO(m)$ and space $\widetilde{O}(n)$ per machine.
    $d_{avg}$ is the average degree of the vertices in the graph.
\end{corollary}

{There is a long line of work on computing approximate triangle counting in
    parallel
    computation~\cite{cohen2009graph,tsourakakis2009doulion,suri2011counting,yoon2011improved,pagh2012colorful,kolountzakis2012efficient,park2013efficient,seshadhri2013triadic,arifuzzaman2013patric,park2014mapreduce,kolda2014counting,jain2017fast, DLSY21}
    and references therein. Despite this progress, and to the best of our
    knowledge, on one hand, each MPC algorithm for exact triangle counting
    either requires strictly super-polynomial in $m$ total space, or the number
    of rounds is super-constant (as seen in~\cref{table:results}).
    On the other hand, the best-known, classic algorithm for
    approximate triangle counting by Pagh and Tsourakakis~\cite{pagh2012colorful}
    requires $T \ge d_{avg}$ even when the space per
    machine is $\Theta(n)$. We design an algorithm that has essentially
optimal total space and round complexity, while at least quadratically improving the requirement on $T$.}

Furthermore, since the amount of messages sent and received by each machine is bounded by $O(n)$,  by~\cite{behnezhad2018semi},  our algorithm directly implies an $O(1)$-rounds algorithm in the \textsc{Congested-Clique} model\footnote{A distributed model where nodes communicate with each other over a complete network using $O(\log n)$ bit messages~\cite{LPPP05}.} under the same restriction $T=\Omega(\sqrt{m/n})$.
The best known (to our knowledge) triangle approximation algorithm for general graphs in this model,
is an ${O}(n^{1/3}/T^{2/3})$-rounds algorithm by~\cite{Dolev12}.
The best-known previous bound only results in constant round complexity when  $T=\Omega(\sqrt{n})$.

\begin{corollary}\label{cor:congest}
    Given a graph $G = (V, E)$ with $T$ triangles, if $T = \Omega(\sqrt{m/n})$, then there exists a $O(1)$-rounds algorithm in the \textsc{Congested-Clique} model that gives a $(1+\eps)$-approximation of $T$ with high probability.
\end{corollary}

The second question we consider is the question of exact counting, for which we  present an algorithm whose total space depends on the arboricity of the input graph.
The arboricity of a graph (roughly) equals the average degree of its densest subgraph.
The class of graphs with bounded arboricity includes many important graph families such as planar graphs,
bounded degree graphs and randomly generated preferential attachment graphs.
In addition, many real-world graphs exhibit bounded arboricity~\cite{GG06,ELS13,shin2018patterns}, making this property important also in practical settings.
For many problems,
a bound on the arboricity of the graph allows for much more efficient algorithms and/or better approximation ratios~\cite{AG09, ELS13}.

Specifically for the task of subgraph counting, in a seminal paper, Chiba and Nishizeki~\cite{CN85} prove that triangle enumeration can be performed in $O(m \alpha)$ time, and assuming 3SUM-hardness this result is optimal up to dependencies in $O(\poly \log n)$~\cite{Patrascu10, KPP16}. Many applied algorithms
also rely on the property of having bounded arboricity in order to achieve better space and time bounds, e.g.,~\cite{ScWa05, chu2011triangle, latapy2008main}.
Our main theorem with respect to this question is the following.
\begin{theorem}\label{thm:exact-tri}
	Let $G=(V,E)$ be a graph with $n$ vertices, $m$ edges and arboricity $\alpha$.
    \textsc{Count-Triangles($G$)} takes $\hideo\left(\log \log
    n\right)$
    rounds,
    $O\left(n^{\delta}\right)$ space per machine for any $\delta > 0$, and $O\left(m\alpha\right)$ total space.
\end{theorem}

It is interesting to note that our total space complexity
matches the time complexity (both upper and conditional lower bounds) of combinatorial\footnote{Combinatorial algorithms, usually, refer to algorithms that do not rely on fast matrix multiplication.}
triangle counting algorithms in the sequential model~\cite{CN85, Patrascu10,KPP16}.
The best-known previous algorithm in this setting is the folklore algorithm of placing each
vertex and its out-neighbors in the same machine and counting the incident triangles. Such
an approach requires $O(\log n)$ rounds and $\Omega(\alpha^2)$ space per machine (summarized
in~\cref{table:results}).
We prove the above theorem in \cref{sec:exact_triangle_counting}.

\subsubsection{Clique Counting}

All of our above triangle counting results can be extended to $k$-clique counting.
In \cref{sec:exact_clique_counting}, we prove that our exact triangle counting result
can be extended to exactly counting $k$-cliques for any constant $k$:
\begin{theorem}\label{thm:exact-clqs}
	Let $G=(V,E)$ be a graph with $n$ vertices, $m$ edges and arboricity $\alpha$.
	\textsc{Count-Cliques($G$)} takes $\hideo\left(\log \log n\right)$ rounds,  $O\left(n^{\delta}\right)$ space per machine
    for any $\delta > 0$, and $O\left(m\alpha^{k-2}\right)$ total space.
\end{theorem}
We can improve on the total space usage if we are given machines where the memory for each individual machine
satisfies $\alpha < n^{\delta'/2}$ where $\delta' < \delta$.
In this case, we obtain an algorithm that counts the number of $k$-cliques in
$G$ using $O(n\alpha^2)$ total space and $\hideo(\log \log n)$ communication rounds.

Furthermore, our approximate triangle counting results can be extended to counting \emph{any}
subgraph of size $K$ where $K$ is constant. Specifically, we obtain the following result:

\begin{theorem}
	Let $G=(V,E)$ be a graph with $n$ vertices, $m$ edges, and let $B$ be the number of
    occurrences of a subgraph $H$ with $K$ vertices in $G$. If $B \geq d_{avg}^{K/2 - 1}$,
    then there exists an MPC algorithm that gives a $(1+\eps)$-approximation of $B$ in
    $O(1)$ rounds, total space $\tO(m)$, and $\tO(n)$ space per machine, with high probability.
    Here, $d_{avg}$ is the average degree of the vertices in the graph.
\end{theorem}

\subsection{Other Small Subgraphs}

Finally, in \cref{sec:5-subgraphs}, we consider the problem of exactly counting subgraphs of size at most $5$, and show that the recent result of Bera, Pashanasangi and Seshadhri~\cite{BPS20} for this question in the sequential model, can be implemented in the MPC model. Ours is the first result for counting any arbitrary subgraph of size at most $5$ in $\poly(\log n)$ rounds in the MPC model.
Here too, our total space complexity matches the time complexity of the sequential model algorithm. It is an interesting open question whether our results can be extended to more general
subgraphs following the results of~\cite{bressan2019faster,bera2021near}. \cref{sec:open} summarizes the difficulties
of implementing these algorithms in the MPC model
and we present this question as interesting future work.

\begin{theorem}
	Let $G=(V,E)$ be a graph with $n$ vertices, $m$ edges, and arboricity $\alpha$.
	The algorithm of BPS for counting the number of occurrences of a subgraph $H$ over $k\leq 5$ vertices  in $G$
	can be implemented in the MPC model
    in $O_{\delta}(\sqrt{\log n}\log \log m)$ rounds, with high probability.
    The space requirement per machine is $O(n^{2\delta})$ and the total space is $O(m\alpha^3)$.
\end{theorem}

%% file: 120-related-work.tex
\subsection{Related Work}
\label{sec:related-work}
There has been a long line of work on small subgraph counting in massive networks in the MapReduce model whose results translate to the MPC
model. We first describe the works for \emph{exact} triangle and $k$-clique counting. \cite{suri2011counting} first designed an algorithm for triangle counting, but their approach requires a super-linear total space of $O(m^{3/2})$. Another work, \cite{afrati2013enumerating}, shows how to count small subgraphs by using $b^3$ machines, each requiring $O(m / b^2)$ space per machine. Hence, it uses a total space of $O(m b)$. Therefore, this approach either requires super-linear total space or almost $O(m)$ space per machine.
\cite{suri2011counting} were the first to achieve constant number of 
rounds in MPC, where they design two
algorithms. The first of those algorithms, that runs in $2$ rounds, requires
$O(\sqrt{m})$ space per machine and total space $O(m^{3/2})$. Their
second algorithm requires only one round for exact triangle counting, total
space $O(\rho m)$ and space per machine $O(m / \rho^2)$. Therefore, for
this algorithm to work with polynomially less than space $m$ per machine, it has
to allow for a total space that is polynomially larger than $m$.
\cite{chu2011triangle} focus on algorithms that require a total space of $O(m)$.
In the worst case, their algorithm performs $O(|E| / S)$ MPC rounds to output
the exact count where $S$ is the maximum space per machine.  \cite{finocchi2015clique} extended and provided new 
algorithms for clique counting but they also require $\Omega(m^{3/2})$ total space.

\cite{tsourakakis2009doulion,arifuzzaman2013patric} designed  randomized
algorithms for \emph{approximate} triangle counting also in the MapReduce model (whose results, again, can be translated rather straightforwardly to the MPC model). 
Their approach first sparsifies the
input graph by sampling a subset of edges, and executes some of the known
algorithms for triangle counting on the sampled subgraph. Denoting their
sampling probability by $p$, their approach outputs a $(1 + \eps)$-approximate triangle
count with probability at most $1 - 1/(\eps^2 p^3 T)$.
\footnote{The actual
probability is even smaller and also depends on pairs of triangles that share an
edge.}
To contrast this result with our approach, consider a graph $G$ where $m
= \Theta(n^2)$. Let $G'$ be the edge-sparsified graph as explained above. To be
able to execute the first algorithm of \cite{suri2011counting} on $G'$ such that
the total space requirement is $O(m)$, one can verify that it is needed to set
$p = \Theta(n^{-2/3})$. This in turn implies that the result in
\cite{tsourakakis2009doulion,arifuzzaman2013patric} outputs the correct
approximation with constant probability only if $T = \Omega(n^2)$. An improved
lower-bound can be obtained by using the second algorithm of
\cite{suri2011counting}. By balancing out $\rho$ and $p$ and for $S =O(n)$,
one can show that the sparsification results in a constant probability of success
for $T = \Omega(n)$.
On the other hand, for $S =O(n)$, our approach obtains the same guarantee
even when $T = \Theta(\sqrt{d_{avg}{(G)}}) = \Theta(\sqrt{n})$.

The best-known algorithm of~\cite{pagh2012colorful} is a randomized algorithm for approximate
triangle counting based on graph partitioning. The graph is
partitioned into $1/p$ pieces, where $p$ is at least the ratio of the maximum
number of triangles sharing an edge and $T$. When all the triangles share one
edge, then $p \ge 1$, and hence such an approach would require the space per
machine to be $\Omega(m)$. Furthermore, this approach requires the number of
triangles to be lower bounded by $T = \Omega\left(d_{avg}\right)$.
Another more recent work of~\cite{seshadhri2013triadic} uses
wedge sampling and provides a $(1+\eps)$-approximation of the
triangle count in $O(1)$ rounds when $T$ is
a constant fraction of the sum of squares of degrees. 
The comparison of our bounds with these previous results are 
summarized in~\cref{table:results}.

\myparagraph{Other related work}
Subgraph counting (primarily triangles) was also extensively studied in the streaming model,  see~\cite{bar2002reductions,kane2012counting,braverman2013hard,jha2013space,mcgregor2016better,bera2017towards,AKK} and references therein. This culminated in a result that requires space $\tO\rb{m^{3/2} / (T \eps^2)}$ to estimate the number of triangles within a $(1 + \eps)$-factor. In the semi-streaming setting it is assumed that one has $\tO(n)$ space at their disposal. This result fits in this regime for $T \ge m^{3/2} / n = d_{avg}\cdot m^{1/2}$. As a reminder, our MPC result requires $T \ge \sqrt{d_{avg}}$ when $S = \tO(n)$.

In a celebrated result, \cite{alon1997finding} designed an
algorithm for triangle counting in the sequential settings
that runs in $O(m^{2 \omega / (\omega +1 )})$ time, where
$\omega$ is the best-known exponent of matrix multiplication.
Since then, several important works have extended this
result to $k$-clique counting~\cite{EG04,V09}.
In the work-depth 
(shared-memory parallel processors) model, several results
are known for this problem. There has been significant work on practical parallel
algorithms for the case of triangle counting (e.g.\ \cite{arifuzzaman2013patric,suri2011counting,park2013efficient,park2014mapreduce,ST15} among others).
There is even an annual competition for parallel triangle counting algorithms~\cite{GraphChallenge}.
For counting $k= 4$ and $k=5$ cliques, efficient practical solutions have
also been developed~\cite{ANRDW17,DAH17,ESBD16,HD14,PSV17}.
\cite{DBS18} recently implemented the Chiba-Nishizeki
algorithm~\cite{CN85} for $k$-cliques
in the parallel setting; although, their work does not achieve polylogarithmic depth.
Even more recently,~\cite{Shi2020} enumerated
$k$-cliques in the work-depth model in $O\left(m\alpha^{k-2}\right)$ expected work and
$O\left(\log^{k-2} n\right)$ depth with high probability,
using $O(m)$ space.
Among other distinctions from our setting, the
work-depth model assumes a shared, common memory.

In the CONGESTED-CLIQUE model, ~\cite{censor2019algebraic} present an $\tilde{O}(n^{1-2/\omega})=\tilde{O}(n^{0.158})$ rounds algorithm for matrix multiplication, implying the same complexity for exact triangle counting. ~\cite{Dolev12} present an algorithm for approximate triangle counting in general graphs whose expected running time is $O(n^{1/3}/T^{2/3})$. They also present an $O(\alpha^2/n)$-rounds algorithm for  bounded arboricity graphs.

%% file: 200-preliminaries.tex
\section{Preliminaries}\label{sec:prelims}

\myparagraph{Counting Duplicates} We make use of interval trees for certain parts of our paper to count the number
of repeating elements in a sorted list, given bounded space per machine.
We use the interval tree implementation
given by~\cite{GSZ11} to obtain our count duplicates algorithm in the MPC model.
We prove the following theorem in the MPC model
regarding our count duplicates tree implementation.
The proofs of the following claims are given in \cref{sec:appendix-prelims}.

\begin{theorem}\label{thm:mpc-interval-tree}
Given a sorted list of $N$ elements implemented on processors where the space per processor is $S$ and the
total space among all processors is $O(N)$, for each unique element in the list,
we can compute the number of times it repeats
in $O\left(\log_{S} N\right)$ communication rounds.
\end{theorem}

We also use the following two new MPC primitives in proving our bounds. These 
primitives may be of use in other algorithms beyond the scope of our paper.

\begin{lemma}\label{lem:queries}
    Given two sets of tuples $Q$ and $C$ (both of which may contain duplicates), for each tuple $q \in Q$,
    we return whether $q \in C$ in $O(|Q \cup C|)$ total space and $\hideo(1)$ rounds given
    machines with space $O(n^{\delta})$ for any $\delta > 0$.
\end{lemma}

\begin{lemma}
\label{lem:duplicate}
    Given a machine $M$ that has space $O(n^{2\delta})$ for any $\delta > 0$ and contains data of $O(n^\delta)$ words,
	we can generate $x$ copies of $M$, each holding the same data as $M$, using $O(M\cdot x)$ machines with $O(n^\delta)$ space each in $O(\log_{n^{\delta}}x )$ rounds.
\end{lemma}

%% file: 220-overview_exact.tex
\subsection{Exact Triangle Counting}
Let $G = (V, E)$ be a graph with $n$ vertices, $m$ edges and
arboricity at most $\alpha$.
We tackle the task of exactly counting the number of triangles in $G$
in $\hideo(\log \log n)$ rounds using the following ideas.
In each round $i$, we partition the vertices into low-degree vertices $A_i$ and high-degree vertices, according to a degree threshold $\gamma_i$,
 which grows doubly exponentially in the number of rounds.  We then count the number of triangles incident to the set of low degree vertices $A_i$.
  Each low-degree vertex $v\in A_i$ sends a list of its neighbors to all its neighbors. Then, any neighbor $u$ of $v$ that detects a common neighbor $w$ to $u$ and $v$,
  adds the triangle $(u,v,w)$ to the list of discovered triangles.

	Once all triangles incident to the vertices in $A_i$ are
  processed, we remove this set from the graph and continue with the
  now smaller graph.
This removal of the already processed vertices
allows us to handle larger and larger degrees from step to step while using a total space of $O(m \alpha)$.
This behavior also leads to the $\hideo(\log \log{n})$ round complexity, as
after this many rounds all vertices are processed.
The key insight in our proof that we maintain $O(m\alpha)$ total space \emph{even when we increase the 
degree threshold doubly exponentially}. Such insight allows us to obtain our improved number of rounds while 
maintaining the same total space as the previous folklore algorithm. Finally, we achieve improved space per 
machine to $O(n^{\delta})$ for any constant $\delta > 0$ via a number of new MPC primitives. Our 
algorithm and its analysis are provided in~\cref{sec:exact_triangle_counting}. We provide extensions of our 
triangle counting algorithm to $k$-cliques in~\cref{sec:exact_clique_counting}.

%% file: 230-overview_estimator.tex
\subsection{Approximate Triangle Counting}
Our work reduces approximate triangle counting to exact triangle counting in multiple (randomly
chosen) induced subgraphs of the original graph.
In our work, and in contrast to prior approaches (e.g., \cite{pagh2012colorful}), the induced sugraphs on different machines might \emph{overlap} in both vertices and edges.
This enables us to obtain better concentration bounds compared to  prior work,
but also brings many challenges. %
Surprisingly, our algorithm is very simple (with a more complicated analysis), but is able to achieve
a better lower bound on the number of triangles required to achieve a $(1+\eps)$-approximation with
high probability.

The high level idea is that each machine $M_i$ samples a subset of vertices  $V_i$ by including each vertex in $V_i$ with probability $\pV$.
Then, each machine computes the induced subgraph $G[V_i]$ and the number of triangles in that subgraph.
The total number of triangles seen across all the machines
is used as an estimator. We repeat in parallel this sampling process
$O(\log n)$ times and return the median of the estimates.
The main challenge this approach raises is: \emph{How do we efficiently collect
\emph{overlapping} induced subgraphs?} (Indeed, approximate triangle counting,
even when the number of triangles is $O(1)$, can be reduced to counting the
number of edges in \emph{sparse} induced subgraphs with the total size of
subgraphs being $\tO(m)$.) We now describe how to handle this task in our
case.

\eat{
\paragraph{Limited space per machine.}
First, we need to make sure that the induced subgraphs do not exceed the space per machine $S$.
This limit implies an upper bound on our sampling probability,
which in turn implies a lower bound on the number of triangles in the graph
from which we can guarantee a good approximation with high probability.
}

\myparagraph{Computing induced overlapping subgraphs}
It is unclear how to compute the induced subgraph on each machine in $O(1)$ rounds without exceeding the total allowed space of $\tilde{O}(m)$.
This task becomes easier if the subgraphs are disjoint. For example, such an issue is avoided when the graph is
\emph{partitioned} across machines as in the algorithm of Pagh and Tsourakakis~\cite{pagh2012colorful}
since there is one copy of each vertex among all the machines. This is not the case for our algorithm.

The trivial strategy of sampling vertices into the machines and querying for all
possible edges between any pair of two vertices takes total space at least
$\sum_{i = 1}^\mM X_i^2$ where $X_i$ is the number of vertices sampled to each
machine $i$. In general, this approach requires much larger than $\tO(m)$ space.
We tackle this challenge by using a \emph{globally known} hash function $h:V\times [\mM] \rightarrow \{0,1\}$,
to indicate whether vertex $v$ is sampled in the $i^{\textrm{th}}$ machine.
By requiring that the  hash function is known to all machines,
we can efficiently compute which edges to send to
each machine, i.e., which edges belong to the subgraph $G[V_i]$.
However, in order for all machines to be able to compute the hash function, the hash function has
to use limited space.
Hence, we cannot hope for a fully independent function, rather we can only use an $(S/\log n)$-wise independent  hash function.
Still, we manage to show that we are able to handle the dependencies introduced by the hash function,
even if we allow as little as $O(\log n)$-independence.

We present our approximate algorithm in~\cref{sec:approx-triangle-counting} and give an extension of this
algorithm to \emph{any} subgraph of size $K$ in~\cref{sec:k-subgraph}.

\eat{
\paragraph{The number of triangles across the machines concentrates.}
Our final challenge is in proving that the sum of triangles over the induced subgraphs is close to its expected value.
Here too we have to take into account the dependencies introduced by the hash function.
}

\eat{
\paragraph{Exploiting the properties of $\MPC$}
The strategy described above takes advantage of  the properties of the MPC model in several ways.
First, we use each individual machine to perform a single
``experiment," and then view the collection of machines
as repeating this experiment $\mM$ times in parallel.
More importantly, our approach exploits the computational power of MPC, even before counting any triangles.
Namely, to compute which machines an edge $e = (u, w)$ should be sent to,
we evaluate the hash function for $u$ and for $w$ w.r.t.~\emph{each} machine.
Then, the intersection of these two lists represent the number of
machines $e$ should be sent to. We show that on average each edge is replicated $O(1)$ times,
while $O(\mM) \gg O(1)$ computation is performed to evaluate the hash function.
This effectively means that we reduce round and communication costs
at the expense of computation. We hope that this technique will find
applications in other problems as well. It is an intriguing question
to obtain a method for induced subgraph partitioning with overlaps whose
computation does not significantly exceed the total size of  the induced subgraphs.
}

%% file: 240-overview_five_subgraphs.tex
\subsection{Counting $k$-cliques and 5-subgraphs.}
We use similar techniques for both problems of exactly counting the number of $k$-cliques and of exactly counting subgraphs up to size $5$. See~\cref{app:exact-k-clique} for details on the former task, and~\cref{sec:5-subgraphs} for details on the latter. Our final result is the first MPC algorithm for counting any \emph{arbitrary} subgraph $H$ of 
size at most $5$ in $\poly(\log n)$ MPC rounds.

Let $H$ denote the subgraph of interest. We say that a subgraph that can be mapped to a subset of $H$ of size $i$ is a \emph{$i$-subcopy of $H$}.
Our main contribution in this section is a new MPC procedure that in each round, tries to extend $i$-subcopies of $H$ to $(i+1)$-subcopies of $H$ by increasing the total space by a factor of at most $\alpha$. This is possible by ordering the vertices in $H$ such that each vertex has at most $O(\alpha)$ outgoing neighbors so that in each iteration only $\alpha$ possible extensions should be considered per each previously discovered subcopy.

\myparagraph{Challenges} The major challenge we face here
is dealing with \emph{finding} and \emph{storing} copies of small (constant-sized)
subgraphs in individual machines. This is a challenge due to the fact that an entire neighborhood of a vertex $v$
may not fit on one machine (recall that we have no restrictions on how large the constant $\delta$ in $O(n^{\delta})$
machine size can be). Thus, we cannot compute all such small subgraphs on one machine. However,
if not done carefully, computing small subgraphs across many machines could potentially result
in many rounds of computation (since we potentially have to try all combinations of vertices
in a neighborhood). We solve this issue by formulating a new MPC procedure
(\cref{lem:duplicate}) in which we
carefully duplicate neighborhoods of vertices across machines.
The detailed analysis of our algorithm is given in~\cref{sec:5-subgraphs}.

%% file: 260-triangle_counting.tex
\section{Exact Triangle Counting in $O(m\alpha)$ Total Space}
\label{sec:exact_triangle_counting}

In this section we describe our algorithm for (exactly) counting the
number of triangles in graphs $G = (V, E)$ of arboricity $\alpha$ and
prove \cref{thm:exact-tri}, restated here, in~\cref{app:exact-triangle-counting-analysis}. We first provide
an overview of our algorithm and its challenges.

\begin{theorem}\label{lem:total-rounds-precise}
	Let $G=(V,E)$ be a graph over $n$ vertices, $m$ edges and arboricity $\alpha$.
	\textsc{Count-Triangles($G$)} takes $\hideo\left(\log \log n\right)$ rounds,  $O\left(n^{\delta}\right)$ space per machine for some constant $0 < \delta < 1$, and $O\left(m\alpha\right)$ total space.
\end{theorem}

Importantly, unlike previous methods, we \emph{do not} need to assume knowledge
of the arboricity of the graph $\alpha$ as input into our algorithm.
The arboricity only shows up in our space bound as a property of the graph but
we do not need to have knowledge of its value as we run the algorithm.
The folklore algorithm shown in~\cref{table:results} requires an assumption of an upper
bound on $\alpha$ since in order to achieve $O(\log n)$ rounds, we must count triangles
incident to and remove all vertices with degree less than or equal to $2\alpha$ in each round.
The procedure gets stuck if we remove vertices with degree $c$ where $c < \alpha$ in each round
because there exists an induced subgraph with degree at least $\alpha$ in a graph with
arboricity $\alpha$.
One can estimate the arboricity of the graph
using $O(\log n)$ additional rounds or an $O(\log n)$ additional factor in space.
Our algorithm does not require this additional step.

In this section, we assume that individual machines have space $\Theta(n^\delta)$
where $\delta$ is some constant $0 < \delta < 1$.
Given this setting, there are several challenges associated with this problem.

\begin{challenge}\label{chal:one-machine}
The entire subgraph neighborhood of a vertex may not fit on a single machine.
This means that all triangles incident to a particular vertex cannot be counted
on one machine. Even if we are considering vertices with degree at most $\alpha$,
it is possible that $\alpha > n^{\delta}$. Thus, we need to have a way
to count triangles efficiently when the neighborhood of a vertex is spread
across multiple machines.
\end{challenge}

The second challenge is to avoid over-counting.

\begin{challenge}\label{chal:duplicates}
When counting triangles across different machines, over-counting the triangles might occur, e.g., if two different machines count the same triangle.
We need some way to deal with duplicate counting of the triangles
to obtain the exact count of the triangles.
\end{challenge}

We deal with the above challenges in our procedures below.
We assume in our algorithm that
each vertex can access its neighbors in $O(1)$ rounds of communication;
such can be ensured via standard MPC techniques.
Let $d_Q(v)$ be the degree of $v$ in the subgraph induced by vertex set $Q$,
i.e.\ in $G[Q]$. Our main algorithm consists of the following
$\textsc{Count-Triangles}(G)$ procedure.

\begin{tboxalg}{{\bf Count-Triangles$(G = (V, E))$}}\label{alg:count}
\begin{algorithmic}[1]

  \State Let $Q_i$ be the set of vertices not yet processed by iteration $i$.
  Initially set $Q_0 \leftarrow V$.
  \State Let $T$ be the current count of triangles. Set $T \leftarrow 0$.

  \For{$i=0$ to $i = \ceil{\log_{3/2}(\log_2(n))}$}
    \State $\gamma_i \leftarrow 2^{(3/2)^i}$.
      \State Let $A_i$ be the list of vertices $v \in Q_i$ where $d_{Q_i}(v) \leq \gamma_i$. Set $Q_{i+1} \leftarrow Q_i \setminus A_i$.
  		\ParFor{$v \in A_i$}
  		 	\State Retrieve the list of neighbors of $v$ and denote it by $L_v$.
  		 	\State\label{line:send-L_v} Send each of $v$'s neighbors a copy of $L_v$.
  		\EndParFor
  		\ParFor{$w \in Q_i$}
  		    \State Let $\mathcal{L}_w = \bigcup_{v \in (N(w) \cap A_i)} L_v$ be
            the union of neighbor lists received by $w$.
  		    \State Set $T \leftarrow T + \textsc{Find-Triangles}(w, \mathcal{L}_w)$.
            \Comment{\cref{alg:find_overlap}}
  		\EndParFor
  \EndFor
  \State Return $T$.
\end{algorithmic}
\end{tboxalg}

\emph{Round compression} is a technique formulated by~\cite{onak18,czumaj2017round} that randomly partitions
the vertices in a graph across machines where each machine then stores the induced subgraph
induced by the partition. Then, a problem (e.g.\ maximum matching) is solved locally in each induced subgraph
in each machine. The solutions in each machine allows one to remove certain vertices, reducing the degree of the
remaining graph.
In each round compression step, the maximum degree of the graph drops by a polynomial factor.
This degree reduction then allows for more aggressive sampling in the next round compression step. This leads
to $O(\log \log \Delta)$ round compression steps until the maximum degree is $\poly(\log n)$; in this
case, the remaining graph can be placed on a single machine.

Our algorithm, although similar, is simpler than the round compression technique. We do not require
sampling since vertices are assigned to machines by degree, deterministically. The crux of our argument
is showing that allowing for total space in terms of the arboricity $\alpha$ leads to a simpler and
deterministic argument. Furthermore, for this specific problem, we also do not need to place the induced
subgraph on one machine. In the next section, we show an implementation that allows us to operate in
the sublinear space per machine regime. We hope our algorithm and analysis will lead to other deterministic
algorithms for bounded arboricity graphs in sublinear space per machine and $O(\log \log n)$ rounds.

\subsection{MPC Implementation Details}

In order to implement $\textsc{Count-Triangles}(G)$ in the MPC model,
we define our $\textsc{Find-Triangles}(w, \mathcal{L})$ procedure
and provide additional details on sending and storing neighbor lists across different machines.
We define \emph{high-degree} vertices
to be the set of vertices whose degree is $> \gamma$ and \emph{low-degree} vertices
to be ones whose degree is $\leq \gamma$ (for some $\gamma$ defined in our algorithm).
We now define the function \textsc{Find-Triangles($w, \mathcal{L}$)} used in the above procedure:

\begin{tboxalg}{{\bf Find-Triangles$(w, \mathcal{L}_w)$}}\label{alg:find_overlap}
\begin{algorithmic}[1]
  \State Sort all elements in $(\mathcal{L}_w \cup (N(w) \cap Q_i))$ lexicographically,  using the procedure given in Lemma 4.3 of~\cite{GSZ11}.
  Let this sorted list of all elements be $S$.
  \State Let $T$ denote the corrected\footnote{ Some care must be taken here to avoid over-counting, since a distinct triangle can show up as several counted duplicates. See~\cref{alg:overcounting} for details.} number of duplicates in $S$ using \cref{thm:mpc-interval-tree}.
  \State Return $T$.
\end{algorithmic}
\end{tboxalg}

\myparagraph{Allocating machines for sorting} Since each $v \in Q_i$
could have multiple neighbors whose degrees are $\leq \gamma$,
the total size of all neighbor lists $v$ receives
could exceed their allowed space $\Theta\left(n^\delta \right)$.
Thus, we allocate $O\left(\frac{\gamma d_{Q_i}(v)}{n^\delta} \right)$
machines for each vertex $v \in Q_i$ to store all neighbor lists that $v$ receives.

The complete analysis for~\cref{lem:total-rounds-precise} is given in~\cref{app:exact-triangle-counting-analysis}.

We provide two additional extensions of our triangle counting algorithm to counting $k$-cliques:

\begin{theorem}\label{thm:k-clique-counting}
    Given a graph $G = (V, E)$ with arboricity $\alpha$, we can count all $k$-cliques in $O(m\alpha^{k-2})$ total
    space, $\hideo(\log \log n)$ rounds, on machines with $O(n^{2\delta})$ space for any $0 < \delta < 1$.
\end{theorem}

We can prove a stronger result when we have some bound on the arboricity of our input graph. Namely,
if $\alpha = O(n^{\delta'/2})$ for any $\delta' < \delta$, then we obtain the following result:

\begin{theorem}\label{thm:query-alg-runtime}
    Given a graph $G = (V, E)$ with arboricity $\alpha$ where $\alpha =
    O(n^{\frac{\delta'}{2}})$ for any $\delta'< \delta$,
we can count all $k$-cliques in $O\left(n\alpha^2\right)$ total space
and $\hideo(\log\log n)$ rounds, on machines with $O(n^{\delta})$ space for any $0 < \delta < 1$.
\end{theorem}

The proofs of these theorems are provided in~\cref{sec:exact_clique_counting}.

%% file: 450-exact-appendix.tex
\subsection{Detailed Analysis}\label{app:exact-triangle-counting-analysis}
In this section we give the full details and analysis of algorithm~\cref{alg:count} given in~\cref{sec:exact_triangle_counting}, for exactly counting the number of triangles in the graph.

We first provide a detailed version of~\cref{alg:find_overlap} that also takes into account over counting due to the fact that each triangle might be counted by several endpoints, and then continue to prove the main theorem of this section,~\cref{thm:exact-tri}.

\input{270-exact-counting-duplicates}

\subsubsection{Proof of~\cref{thm:exact-tri}}
First, all proofs below assume we start at a cutoff of $\gamma =
4\alpha$. Because we increase the cutoff bound doubly exponentially, we can
reach such a bound in $O(\log \log \alpha)$ rounds. Thus, in the following
proofs, we ignore all rounds before we get to a round where $\gamma \geq
4\alpha$.
Before proving the theorem, we provide several useful lemmas stating that the number of vertices and edges remaining at the beginning of each iteration is bounded.
\begin{lemma}\label{lem:vertex-pruning}
At the beginning of iteration $i$ of $\textsc{Count-Triangles}$, given $\gamma_i = 2^{(3/2)^{i}}\cdot (2\alpha)$ as stated in~\cref{alg:count},
the number of remaining vertices $N_i = |Q_i|$
is at most $\frac{n}{2^{2\cdot ((3/2)^i - 1)}}$.
\end{lemma}
\begin{proof}
Let $N_i$ be the number of vertices in $Q_i$ at the beginning of iteration $i$.
Since the subgraph induced by $Q_i$ must have arboricity bounded by $\alpha$, we can bound the total degree of $Q_i$,
\[
\mathlarger\sum\limits_{v\in Q_i} d_{Q_i}(v) < 2\alpha |Q_i| = 2N_i\alpha.
\]

At the end of the iteration, we only keep the vertices in $Q_{i+1} = \{ v\in Q_i \mid d_{Q_i}(v) > \gamma_i\}$.
If we assume that $|Q_{i+1}| > \frac{N_i}{\gamma_i/(2\alpha)}$, then we obtain a contradiction since this implies that
\[
\mathlarger\sum\limits_{v\in Q_{i+1}} d_{Q_i}(v) > |Q_{i+1}|\cdot\gamma_i > 2N_i\alpha > \mathlarger\sum\limits_{v\in Q_i} d_{Q_i}(v).
\]

Then, the number of remaining vertices follows directly from the above by induction on $i$ with base case $N_1 = n$,
\[
N_i \le  \frac{N_{i-1}}{\gamma_i/(2\alpha)} = \frac{N_{i-1}}{2^{(3/2)^{i-1}}} \leq
\frac{n}{\prod\limits_{j=0}^{i-1} 2^{(3/2)^j}} = \frac{n}{2^{2\cdot \left( (3/2)^i - 1 \right)}}.
\]
\end{proof}

We can show a similar statement for the number of edges that remain at the start of the $i^{th}$ iteration.

\begin{lemma}\label{lem:edge-pruning}
    At the beginning of iteration $i$ of $\textsc{Count-Triangles}$, given $\gamma_i$,
    the number of remaining edges $m_i$ is at most $m_i \leq \frac{m}{2^{2\cdot((3/2)^{i-1} - 1)}}$.
\end{lemma}

\begin{proof}
    The number of vertices remaining at the
    beginning of iteration $i$ is given by $|Q_i|$.
    Thus, because the arboricity of our graph is $\alpha$, we can upper bound
    $m_i$ by
    \[
        m_i \leq |Q_i|\alpha.
    \]

    Then, we can also lower bound the number of edges at the beginning of iteration $i - 1$
    since the vertices that remain at the beginning of round $i$
    are ones which have greater than $\gamma_{i-1}$ degree,
    \[
        m_{i-1} \geq \frac{1}{2}\sum_{v \in Q_{i-1}}d_{Q_{i-1}}(v) \geq \frac{1}{2}|Q_i|\gamma_{i-1}.
    \]

    Thus, we conclude that $m_i \leq \frac{2\alpha m_{i-1}}{\gamma_{i-1}}$. By induction on $i$ with
    base case $m_0 = m$, we obtain,
    \[
        m_i \leq 2\alpha \left(\frac{m_{i-1}}{\gamma_{i-1}}\right) \leq  \frac{m}{\prod_{j = 0}^{i-2}2^{(3/2)^j}} =\frac{m}{2^{2\cdot((3/2)^{i-1} - 1)}}.
    \]
\end{proof}

The above lemmas allows us to bound the total space used by the algorithm.
\begin{lemma}\label{lem:total-space}
\textsc{Count-Triangles($G$)} uses $O(m\alpha)$ total space when run on a graph $G$ with arboricity $\alpha$.
\end{lemma}

\begin{proof}
The total space the algorithm requires is the sum of the
space necessary for storing the neighbor lists sent by
all vertices with degree $\leq \gamma_i$ and the space necessary for all vertices to store their own neighbor lists.
The total space necessary for each vertex to store its own neighbor list is $O(m)$.

Now we compute the total space used by the algorithm during iteration $i$.
The number of vertices in $Q_i$ at the beginning of this iteration
is at most $N_i \leq \frac{n}{2^{2\cdot \left( (3/2)^i - 1 \right)}}$ by~\cref{lem:vertex-pruning}.
Each vertex $v$ with $d_{Q_i}(v) \le \gamma_i$, makes
$d_{Q_i}(v)$ copies of its neighbor list ($N(v) \cap Q_i$)
and sends each neighbor in $N(v) \cap Q_i$ a copy of the list.
Thus, the total space required by the messages sent by $v$ is $d_{Q_i}(v)^2 \le \gamma_i^2$.
$v$ sends at most one message of size $d_{Q_i}(v) \le \gamma_i$ along
each edge $(v, w)$ for $w \in N(v) \cap Q_i$.
Then, by \cref{lem:edge-pruning}
the total space required by all the low-degree vertices in round $i$ is at most (as at most two messages are sent along
each edge):
\[
    2m_i\cdot \gamma_i < \frac{m}{2^{2\cdot \left( (3/2)^{i-1} - 1 \right)}}\cdot \left[ 2^{(3/2)^i} (2\alpha)\right] = 16 m\alpha.
\]
\end{proof}

We are now ready  to prove~\cref{thm:exact-tri}.
\begin{proof}[Proof of \cref{thm:exact-tri}]
By~\cref{lem:vertex-pruning}, the number of vertices remaining in $Q_i$ at the beginning of iteration $i$
is $\frac{n}{2^{2\cdot \left( (3/2)^i - 1 \right)}}$. This means that the procedure runs for $O(\log \log n)$ iterations
before there will be no vertices. For each of the $O(\log \log n)$ iterations,
\textsc{Count-Triangles($G$)} uses $\hideo(1)$ rounds of communication
for the low-degree vertices to send their neighbor lists to their neighbors. The algorithm
then calls \textsc{Find-Triangles-Exact($w, \mathcal{L}_w$)} on each vertex $w \in Q_i$ (in parallel)
to find the number of triangles incident to $w$ and vertices in $A_i \subseteq Q_i$. \textsc{Find-Triangles-Exact($w, \mathcal{L}_w$)} requires
$O\left(\log_{n^{\delta}} (m \alpha)\right) = O(1/\delta)$ rounds by Lemma 4.3 of~\cite{GSZ11} and~\cref{thm:mpc-interval-tree}.
Therefore, the total number of rounds required by \textsc{Count-Triangles($G$)}
is $O\left(\frac{\log \log n}{\delta}\right) = \hideo(\log \log n)$.
\end{proof}

\section{Extensions to Exact $k$-Clique Counting in Graphs with Arboricity $\alpha$}\label{sec:exact_clique_counting}

In this section, we briefly provide two algorithms
for exact counting of $k$-cliques (where $k$ is constant) in
graphs with arboricity $\alpha$. The first is
an extension of our exact triangle counting result given in~\cref{sec:exact_triangle_counting}.
The second is a query-based algorithm where the neighborbood of a low-degree vertex is
constructed on a single machine via edge queries. In this case, the triangles incident
to any given low-degree vertex can be counted on the same machine.

\subsection{Exact $k$-Clique Counting}\label{app:exact-k-clique}

\paragraph{Exact $k$-Clique Counting in $O\left(m\alpha^{k-2}\right)$ Total Space and $\hideo(\log \log n)$ Rounds}

We extend our algorithm given in~\cref{sec:exact_triangle_counting} to exactly count $k$-cliques (where
$k$ is constant) in $O\left(m\alpha^{k-2}\right)$ total space and $\hideo(\log \log n)$ rounds. Given
a graph $G = (V, E)$ with arboricity $\alpha$, the idea behind the
algorithm is the following: let $G_i = \left(V_i\cup V, E_k\cup E\right)$ be a  graph where each vertex $v\in V_i$ corresponds  to an $i$-clique in $G$. Let $K(u)$ denote the $K_i \in G$ represented
by $u \in V_k$.
An edge $(u,v)$ exists in $E_i$ iff $u\in V_i$, $v \in V$ and $K(u) \cup \{v\}$ is an $(i+1)$-clique in $G$.
We construct the $G_k$ graphs iteratively, starting with
$G_1 = G$.
Then, given $G_{i-1}$, we recursively
construct $G_i$ by using our exact triangle counting algorithm. Once we have $G_{k-2}$,
we obtain our final count of the number of $k$-cliques by running our exact triangle counting algorithm one last
time. The total space used is dominated by running the triangle counting algorithm on $G_{k-2}$, which uses
$O\left(m\alpha^{k-2}\right)$ total space.
Since we run the triangle counting algorithm $O(k)$ times and $k$ is a constant,
the total number of rounds of communication necessary is $\hideo(\log \log n)$ rounds.
This detailed algorithm
is given below.

Below, we describe our $O(n\alpha^{k-1})$ total space, $O(\log \log n)$ rounds
exact $k$-clique counting algorithm that can be run on machines with space $O(n^{\delta})$.
Calling $\textsc{Count-$k$-Cliques}(G, k, k)$ for any given graph $G = (V, E)$
returns the number of $k$-cliques in $G$.

\begin{tboxalg}{{\bf $k$-Clique-Counting$(G = (V, E), k, k')$}}\label{alg:count-cliques_unimportant}
\begin{algorithmic}[1]
  \If{$k \leq 1$}
    \State Return $(|N|, G)$
  \Else
  \State $(x, G_{k-1}) \leftarrow$ \textsc{Count-$k$-Cliques}($G, k-1, k'$)
  \State $T \leftarrow \textsc{Enumerate-Triangles}(G_{k-1})$. Let $T$ be the set of all enumerated triangles.
  \State Initialize sets $V_k \leftarrow \emptyset$ and $E_k \leftarrow \emptyset$.
  \ParFor{$t \in T$}
    \State Let $K(t)$ represent the set of vertices in $V$ composing the clique represented
    by $t \in T$.
    \ParFor{$v \in K(t)$}
    \State Let $v'(S)$ be a vertex $v$ representing a set of vertices $S$.
    In other words, $K(v) = K(v'(S)) = S$.
    \State $V_{k} \leftarrow V_k \cup v'(K(t)\setminus v)$.
    \State $E_k \leftarrow E_k \cup (v, v'(K(t)\setminus v))$.
    \EndParFor
  \EndParFor
  \If{$k = k' - 2$}
    \State Return $|T|$.
  \Else
     \State Return $(|V_k|, G_k(V \cup V_k, E \cup E_k))$.
  \EndIf
\EndIf
\end{algorithmic}
\end{tboxalg}

\begin{tboxalg}{{\bf Triangle-Enumeration$(G=(V,E))$}}\label{alg:triangle-enum}
    \begin{algorithmic}[1]
            \State Let the set of enumerated triangles to be $T \leftarrow \emptyset$.
            \State Let $Q_i$ be the set of vertices that have not yet been processed by iteration $i$.
            Initially set $Q_0 \leftarrow V$.

            \ParFor{$i=0$ to $i = \ceil{\log_{3/2}(\log_2(n))}$}
                \State $\gamma_i \leftarrow 2^{(3/2)^i}\cdot 2\alpha$.
                \State Let $A_i$ be the list of vertices $v \in Q_i$ where $d_{Q_i}(v) \leq \gamma_i$. Set $Q_{i+1} \leftarrow Q_i \setminus A_i$.
                \State Use~\cref{lem:enum-triangle} to enumerate the set of triangles incident to $A_i$. Let this set be $T_i$.
                \State $T \leftarrow T \cup T_i$.
            \EndParFor
        \State Return $T$.
    \end{algorithmic}
\end{tboxalg}

\subsection{MPC Implementation}\label{sec:mpc-k-clique-impl}

To implement $\textsc{Count-$k$-Cliques}$ in the MPC model, we must
be able to create the graph $G_2, \dots, G_{k-1}$ efficiently in our given
space and rounds. The crux of this algorithm is the procedure for enumerating
all triangles given a set $A$ of vertices in $G$ where $d(v) \leq \gamma$ for
all $v \in A$.
To do the triangle enumeration, we prove \cref{lem:enum-triangle} which
can enumerate all such triangles incident to $A$ in $O(m\gamma)$ total
space, $\hideo(1)$ rounds given machines with space $O\left(n^{2\delta}\right)$.

\begin{lemma}\label{lem:enum-triangle}
Given a graph $G$, a constant integer $k\geq 2$, and a subset $A\subseteq G$
of vertices such that for every $v\in A$, $d(v)\leq \gamma$,
we can generate all triangles in $G$ that are incident to
vertices in $A$ in $\hideo(1)$ rounds, $O(n^{2\delta})$ space per machine, and $O(m\gamma)$ total space.
\end{lemma}
\begin{proof}
Let $R$ be the set of machines holding the edges incident to $A$.
Here too, similarly to the proof of~\cref{lem:DRTS}, it will be easier to think of
each machine $M$ as a set of $n^{\delta}$ parts, so that
each edge, incident to a vertex in $A$, resides on a single part.
We duplicate each such  part, holding some neighbor of $A$, $\alpha$ times, using~\cref{lem:duplicate}.
(We will actually duplicate machines, but, again, think of the duplicated machines as a collection of duplicated parts.)
By~\cref{lem:duplicate}, this takes $O(\log_{n^{\delta}}\alpha)=\hideo(1)$ rounds.
Fix some vertex $v\in A$ and assume that  $u\in N(v)$ resides on part $P_i(v)$.
After the duplication step, there are $\alpha$ copies of each part. We denote
these copies $P_{i, 1}(v), \dots, P_{i, \alpha}(v)$.
All parts $P_{i, j}(v)$ where $j \in [\alpha]$ and $v \in A$ then asks for $v$'s $i$-th neighbor
in $O(1)$ rounds of communication.
Now, each part $P_{i, j}(v)$ creates $O(1)$ edge queries to check whether its vertices form a triangle.
All of the queries generated by all parts can be answered in parallel using~\cref{lem:queries} in $\hideo(1)$ rounds.
Then each part that discovered a triangle incident to $v$ adds it to a list $K$.
Now we sort the list $K$ and remove any duplicated triangles,
so that the list only holds a single copy of every clique incident to some vertex in $A$.
The total round complexity is $\hideo(1)$ due to the duplications, sorting, and answering the queries.
The space per machine is $O(n^{2\delta})$ and the total memory is $O(m\alpha)$
as each machine was duplicated $\alpha$ times.
\end{proof}

Using~\cref{lem:enum-triangle}, we can now prove the space usage and round complexity of $\textsc{Enumerate-Triangles}$.

\begin{lemma}\label{lem:enum-triangles}
    Given a graph $G = (V, E)$ with arboricity $\alpha$, $\textsc{Enumerate-Triangles}(G)$ uses
    $O(n\alpha^2)$ total space, $\hideo(\log \log n)$ rounds on machines with $O(n^{2\delta})$
    space.
\end{lemma}

\begin{proof}
    By~\cref{lem:vertex-pruning}, the number of vertices remaining in $Q_i$ at the beginning of the
    $i$-th iteration of $\textsc{Enumerate-Triangles}$ is at most $\frac{n}{2^{2\cdot \left((3/2)^i-1\right)}}$.
    By~\cref{lem:enum-triangle}, the total space usage of enumerating all triangles incident to $A_i$
    is $O(m\gamma_i) = O\left(m \cdot \left(2^{(3/2)^i} \cdot 2\alpha\right)\right)$. The
    summation of the space used for all $i$ is then:
    \begin{align*}
        \sum_{i = 0}^{\ceil{\log_{3/2}(\log_2(n))}} \left(\frac{n}{2^{2\cdot \left((3/2)^i-1\right)}}\right) \cdot \left(2^{(3/2)^i} \cdot 2\alpha^2\right) = O(n\alpha^2).
    \end{align*}

    The number of rounds required by this algorithm is $O(\log \log n) \cdot \hideo(1) = \hideo(\log \log n)$.
\end{proof}

Given the total space usage and number of rounds required by $\textsc{Enumerate-Triangles}$, we can
now prove the total space usage and number of rounds required by $\textsc{Count-$k$-Cliques}$.
But first, we show that for any graph $G = (V, E)$ with arboricity $\alpha$, all graphs
$G_1, \dots, G_{k-1}$ created by $\textsc{Count-$k$-Cliques}$ has arboricity $O(\alpha)$
for constant $k$.

\begin{lemma}\label{lem:clique-graph-arboricity}
    Given a graph $G = (V, E)$ with arboricity $\alpha$ as input to $\textsc{Enumerate-Triangles}$,
    all graphs $G_1, \dots, G_{k-1}$ generated by the procedure have arboricity $O(\alpha)$
    for constant $k$.
\end{lemma}

\begin{proof}
    We prove this lemma via induction. In the base case, $G_1 = G$ and so $G_1$ has
    arboricity $\alpha$. Now we assume that $G_i$ for $i \in [k-1]$ has arboricity $O(\alpha)$
    (for constant $i$) and show that $G_{i+1}$ has $O(\alpha)$ arboricity.
    Suppose that $G_{i}$ has arboricity $c\alpha$ for some constant $c$.
    We prove via contradiction that the arboricity of $G_{i+1}$ is upper bounded
    by $3(i+1)c\alpha$. Suppose for the sake of contradiction that the arboricity of $G_{i+1}$
    is greater than $3(i+1)c\alpha$. Then, there must exist a subgraph,
    $G_{i+1}[V']$ for some vertex set, $V'$, of $G_{i+1}$ that
    contains greater than $3(i+1)c\alpha|V'|$ edges (by definition of arboricity).
    We now convert this subgraph $G_{i+1}[V']$ to a subgraph in $G_{i}$.
    Every vertex in $V'$ maps to at most $i$ pairs of vertices in $G_{i}$
    connected by an edge. Every edge in $G_{i+1}[V']$ maps to at least
    $1$ edge. Thus, the subgraph in $G_{i}$ that $G_{i+1}[V']$ maps to
    contains at most $2i|V'|$ vertices and at least $3(i+1)c\alpha|V'|$ edges.
    This implies, by the definition of arboricity, that the arboricity of $G_i$
    is $\geq \frac{3(i+1)c\alpha|V'|}{2i|V'|} > c\alpha$, a contradiction.
    Hence, the arboricity of $G_{i+1}$ is at most $3(i+1)c\alpha$. And we have
    proven that the arboricity of $G_{i+1}$ is $O(\alpha)$ for constant $k$.
    By induction, all graphs $G_1, \dots, G_{k-1}$ have arboricity $O(\alpha)$.
\end{proof}

Now we prove our final theorem of the space and round complexity of $\textsc{Count-$k$-Cliques}$.

\begin{proof}[Proof of~\cref{thm:k-clique-counting}]
    The number of $i$-cliques in a graph with arboricity $\alpha$ is at most
    $O(m\alpha^{i-2})$. Thus, by~\cref{lem:enum-triangles} and~\cref{lem:clique-graph-arboricity},
    $\textsc{Count-$k$-Cliques}$ during the $i$-th call uses $O(m\alpha^{i})$ total space,
    $\hideo(\log \log n)$ rounds. Thus, $\textsc{Count-$k$-Cliques}$ uses $O(m\alpha^{k-2})$ space,
    $\hideo(\log \log n)$ rounds given machines with $O(n^{2\delta})$ space to count $k$-cliques given
    that the procedure terminates on the $(k-2)$-th iteration.
\end{proof}

\subsection{Exact $k$-Clique Counting in $O\left(n\alpha^2\right)$ Total Space and $\hideo(\log \log n)$ Rounds}
We can improve on the total space usage if we are given machines where the memory for each individual machine
satisfies $\alpha < n^{\delta'/2}$ where $\delta' < \delta$.
In this case, we obtain an algorithm that counts the number of $k$-cliques in
$G$ using $O(n\alpha^2)$ total space and $\hideo(\log \log n)$ communication rounds.

The entire neighborhood of any vertex with degree $\leq n^{\delta/2}$
can fit on one machine. Suppose that $\alpha < n^{\delta'/2}$ where $\delta' < \delta$,
then, there will always exist vertices that have degree $\leq n^{\delta/2}$. Our algorithm
proceeds as follows:

\begin{tboxalg}{{\bf Count-Cliques$(G = (V, E))$}}\label{alg:count-cliques}
\begin{algorithmic}[1]
  \State Let $Q_i$ be the set of vertices that have not yet been processed by iteration $i$. Initially set
  $Q_0 \leftarrow V$.
  \State Let $C$ be the current count of cliques. Set $C \leftarrow 0$.
  \For{$i = 0$ to $i = \ceil{\log_{3/2}(\log_2(n))}$}
    \State $\gamma_i \leftarrow 2^{(3/2)^i} \cdot 2\alpha$.
    \State Let $A_i$ be the list of vertices where $d_{Q_i}(v) \leq \min(cn^{\delta/2}, \gamma_i)$ for some constant $c$.
    \State Set $Q_{i+1} \leftarrow Q_i \setminus A_i$.
    \ParFor{$v \in A_i$}
        \State Retrieve all neighbors of $v$. Let this list of $v$'s neighbors be $L_v$.
        \State Query for all pairs $u, v \in L_v$ to determine whether edge $(u, v)$ exist. Retrieve
        all edges that exist.
        \State Count the number of triangles $T_v$ incident to $v$, accounting for duplicates.
        \State $T \leftarrow T + T_v$.
    \EndParFor
  \EndFor
\end{algorithmic}
\end{tboxalg}

\subsection{MPC Implementation Details}

\paragraph{Accounting for Duplicates} We account for duplicates by counting for each iteration $i$ how
many triangles on each machine contains $1$, $2$ or $3$ vertices which have degree $\leq \min(cn^{\delta/2}, \gamma_i)$
(again we call these vertices low-degree).
We multiply the count of triangles which have $t \geq 2$ low-degree vertices by $\frac{1}{t}$ to correct
for over-counting due to multiple low-degree vertices
performing the count on the same triangle. Each machine can retrieve
the degrees of vertices in it in $\hideo(1)$ rounds and such information can be stored on the machine
given sufficiently small constant $c$ in $\textsc{Count-Clique}$.

\begin{proof}[Proof of~\cref{thm:query-alg-runtime}]
Since we are considering vertices with degree at most $\min(cn^{\delta/2}, \gamma_i)$,
by \cref{lem:total-space}, the total space used by our algorithm during any iteration $i$ is
\begin{align*}
    N_i \cdot \left(\min\left(cn^{\delta/2}, \gamma_i \right)\right)^2 < 16n\alpha^2.
\end{align*}

By~\cref{lem:queries}, we query for whether each of the $\min\left(cn^{\delta/2}, \gamma_i\right)^2$
potential edges on each machine is an edge in $G$ in parallel using $O(n\alpha^2)$ total
space and $\hideo(1)$ rounds.

If $\gamma_i < cn^{\delta/2}$ for all iterations $i$,
then by \cref{lem:total-rounds-precise}, the number of communication rounds
required by $\textsc{Count-Cliques}$ is $\hideo(\log \log n)$. If, on the other hand,
$cn^{\delta/2} < \gamma_i$, then the number of vertices remaining in $Q_i$ decreases
by a factor of $cn^{\delta/2}$ every round. Thus, the number of rounds required in this case is
$O\left(\frac{2+\delta'}{\delta}\right)$. Since we assume $\delta'$ and $\delta$ are constants,
the number of communication rounds needed by this algorithm is $\hideo(\log \log n)$.
\end{proof}

%% file: 270-exact-counting-duplicates.tex
\subsubsection{Details about finding duplicate elements using \cref{thm:mpc-interval-tree}}\label{sec:overcounting}
$\textsc{Find-Triangles}(w, \mathcal{L}_w)$ finds triangles by counting the number of duplicates that
occur between elements in lists.
~\cref{thm:mpc-interval-tree} provides a MPC implementation for
finding the count of all occurrences of every element in a sorted list.
Provided a sorted list of neighbors of $v \in Q_i$ and neighbor lists in $\mathcal{L}_v$,
this function counts the number of intersections between a neighbor list sent to $v$ and the neighbors of $v$.
Every intersection indicates the existence of a triangle. As given, $\textsc{Find-Triangles}(w, \mathcal{L}_w)$ (see v~\cref{alg:find_overlap}) 
returns a $6$-approximation of the number of triangles in any graph. We provide a detailed and somewhat more complicated
algorithm $\textsc{Find-Triangles-Exact}(w, \mathcal{L}_w)$ that accounts for over-counting of triangles and returns the
exact number of triangles.

Since~\cref{thm:mpc-interval-tree} returns the total count of
each element, we subtract the value returned by $1$ to obtain the number of intersections.
Finally, each triangle containing one low-degree vertex will be counted twice, each containing two low-degree vertices
will be counted $4$ times, and each containing three low-degree vertices will be counted $6$ times. Thus,
we need to divide the counts by $2$, $4$, and $6$, respectively, to obtain the exact count of unique triangles.

\begin{tboxalg}{{\bf Find-Triangles-Exact$(w, \mathcal{L}_w)$}}\label{alg:overcounting}
\begin{algorithmic}[1]
  \State Set the number of triangles $T_i \leftarrow 0$.
  \State Sort all elements in $(\mathcal{L}_w \cup (N(w)\cap Q_i))$ lexicographically  using the procedure given in Lemma 4.3 of~\cite{GSZ11}.
  Let this sorted list of all elements be $S$.
  \State Count the duplicates in $S$ using \cref{thm:mpc-interval-tree}.
  \ParFor{all $v \in N(w)$}
    \State Let $R$ be the number of duplicates of $v$ returned by \cref{thm:mpc-interval-tree}.
    \If{$d_{Q_i}(v) > \gamma_i$ and $d_{Q_i}(w) > \gamma_i$}
        \State Increment $T_i \leftarrow T_i + \frac{R-1}{2}$.
    \ElsIf{($d_{Q_i}(v) > \gamma_i$ and $d_{Q_i}(w) \leq \gamma_i$)
    or ($d_{Q_i}(v) \leq \gamma_i$ and $d_{Q_i}(w) > \gamma_i$)}
        \State Increment $T_i \leftarrow T_i + \frac{R-1}{4}$.
    \Else
   		\State Increment $T_i \leftarrow T_i + \frac{R-1}{6}$.
   \EndIf
   \EndParFor
  \State Return $T_i$.
\end{algorithmic}
\end{tboxalg}

Substituting $\textsc{Find-Triangles-Exact}$ in $\textsc{Count-Triangles}$ finds the exact count of
triangles in graphs with arboricity $\alpha$ using $O(m\alpha)$ total space.

%% file: 300-triangle-estimates.tex
\section{Approximate Triangle Counting in General Graphs}
\label{sec:approx-triangle-counting}

In this section we provide our algorithm for estimating the number of triangles in general graphs
(see \cref{alg:approx-count-triangles,alg:approx-count-triangles-main}) and hence prove \cref{thm:approximate-counting}.
\EstimatorGrand*
The rationale behind the lower bound constraints in \cref{thm:approximate-counting} will become clear when we discuss the challenges
and analysis (formally presented in the following sections). %

%% file: 305-estimator_challenges.tex
\subsection{Overview of the Algorithm and Challenges}
\label{sec:overview_of_the_algorithm_and_challenges}
Our approach is to use the collection of machines to repeat the following experiment multiple times in parallel.
Each machine $M_i$ samples a subset of vertices $V_i$,
and then counts the number of triangles $\hat T_i$ seen in each induced graph $G[V_i]$.
We then use the sum $\hat T$ of all $\hat T_i$'s  as an unbiased estimator (after appropriate scaling) for the number of triangles $T$ in the original graph.

\begin{tboxalg}{{\bf Approximate-Triangle-Counting(G=(V,E))}}\label{alg:approx-count-triangles}
\begin{algorithmic}[1]
    \State $R\gets 0$
    \ParFor{$i \gets 1 \ldots \mM$}

        \State{Let $V_i$ be a random subset of $V$}
            \Comment{See \cref{sec:ensuring_that_g_v_i_fits_on_a_single_machine} for  details about the sampling} \label{line:vertex_subset}

        \If{size of $G[V_i]$ exceeds machine space $S$} \label{line:check-size-of-G[V_i]}
            \State Ignore this sample and set $\hat T_i\gets 0$
        \Else
            \State Let $\hat T_i$ be the number of triangles in $G[V_i]$ \label{line:approx-count-define-Xi}
            \State $R \gets R+1$
        \EndIf

    \EndParFor
    \State Let $\hat T = \sum_{i = 1}^{\mM} \hat T_i$  \label{line:approx-count-define-X}
    \State \Return  $\frac{1}{\pV^3 R} \hat T$ \label{line:approx-count-return}
\end{algorithmic}
\end{tboxalg}

Moving forwards, for the most part, we will focus on a specific machine $M_i$ containing $V_i$ (a single experiment).
We list the main challenges in the analysis of this algorithm, along with the
sections that describe them.
\begin{enumerate}
    \item \textbf{\cref{sec:ensuring_that_g_v_i_fits_on_a_single_machine}:}
    The induced subgraph $G[V_i]$ fits into the memory $S$ of $M_i$
    (thus allowing us to count the number of triangles in  $G[V_i]$ in one round).
    \label{ichal:first_challenge}
    \item \textbf{\cref{sec:computing_the_induced_subgraph}:}
    We can efficiently (in one round)  collect all the edges in the induced subgraph $G[V_i]$.
    This involves presenting an $\MPC$ protocol such that
    the number of messages \emph{sent and received} by any machine is at most the \emph{space per machine} $S$.
    \label{ichal:second_challenge}
    \item
    \textbf{\cref{sec:sub_challenge_showing_that_protocol_succeeds}} With high constant probability, the number of messages sent and received by each machine $M_i$ is at most $S$.    \label{ichal:bound_msgs_challenge}
    \item \textbf{\cref{sec:estimating_number_of_triangles_from_the_aggregate}:}
        With high \emph{constant} probability (of at least $0.9$), 
        the sum of triangles across all machines, $\hat T$, is close to its expected value. Then, repeating the algorithm polylogarithmic number of times
        with only a polylogarithmic increase in total space,
        and by using the median trick, allows us to get a high probability bound.
        The specifics are discussed in~\cref{sec:getting_the_high_probability_bound}.
    \label{ichal:third_challenge}
\end{enumerate}

In each of the following sections, we first present a high level overview of the challenges that we need
to solve and then follow these high-level descriptions with detailed proofs.

%% file: 310-fit_on_machine.tex
\subsection{Challenge~(\ref{ichal:first_challenge}): Ensuring That $G[V_i]$ Fits on a Single Machine}
\label{sec:ensuring_that_g_v_i_fits_on_a_single_machine}

\paragraph*{Ensuring that \emph{edges} fit on a machine:}
\label{par:ensuring_that_edges_fit_on_a_machine}
Our algorithm constructs $V_i$ by including each $v\in V$ with probability $\pV$,
which implies that the expected number of edges in $G[V_i]$ is $\pV^2 m$.
Since we have to ensure that each induced subgraph $G[V_i]$ fits on a single machine, we obtain the constraint $\pV^2 m = O(S)$.
Concretely, we achieve this by defining:
\begin{equation}\label{eq:define-pV}
\pV \eqdef \frac{1}{10}\cdot\sqrt{\frac{S}{mk}}\;,
\end{equation}
where the parameter $k = O(\log n)$ will be exactly determined later (See \cref{sec:computing_the_induced_subgraph}).

\paragraph*{Ensuring that \emph{vertices} fit on a machine:}
\label{par:ensuring_that_vertices_fit_on_a_machine}
In certain regimes of values of $n$ and $m$, the \emph{expected number of vertices} ending up in an induced subgraph --
$\pV n$, may exceed the space limit $S$.
Avoiding this scenario introduces an additional constraint $\pV n = O(S) \iff S = \Omega(kn^2/m)$.

\paragraph*{Getting a high probability guarantee:}
\label{par:getting_a_high_probability_guarantee}
As discussed above, the value of $\pV = \widetilde\Theta_{\eps}(\sqrt{S/m})$ is chosen specifically so that
the \emph{expected number of edges} in the induced subgraphs $G[V_i]$ is $\pV^2 m\le\Theta(S)$, thus using all the available space (asymptotically).
In order to guarantee that this bound holds \emph{with high probability} (see \cref{sec:light_vertices}),
we require additional constraints on the space per machine $S = \widetilde{\Omega}_{\eps}(\sqrt{m})$.
We remark that this lower bound $S = \widetilde{\Omega}_{\eps}(\sqrt{m})$ is essentially saying that $\mM = \widetilde O_{\eps}(\sqrt{m})$,
i.e. the \emph{space per machine} is much larger than the \emph{number of machines}.
This is a realistic assumption as  in practice we can have machines with $10^{11}$ words of local random access memory,
however, it is unlikely that we also have as many machines in our cluster.

\paragraph*{Lower Bound on \emph{space per machine}:}
\label{par:lower_bound_on_memory_per_machine}
Combining the above two constraints, we get:
\begin{align}
\label{eq:constraints_on_space_per_machine}
S > \max\left\{15\frac{\sqrt{mk}}{\eps}, \frac{100kn^2}{m} \right\} \implies
S = \widetilde{\Omega}_{\eps}\left( \max\left\{\sqrt{m}, \frac{n^2}{m} \right\} \right)
\end{align}
Note that \cref{eq:constraints_on_space_per_machine} always allows \emph{linear space per machine}, as long as $m = \Omega(n)$.
The following sections,
\cref{sec:bounding_the_number_of_light_edges_on_a_machine,,sec:bounding_the_number_of_heavy_edges_on_a_machine} present a detailed analysis,
showing that the number of vertices and edges in each subgraph is at most $S$ \emph{with high probability}. In this high-level overview of the challenges, we defer a detailed analysis 
of these bounds to the later sections (\cref{sec:bounding_the_number_of_light_edges_on_a_machine,,sec:bounding_the_number_of_heavy_edges_on_a_machine}) since the formal proof of these bounds also require a discussion of~\cref{sec:computing_the_induced_subgraph}.

%% file: 315-induced_subgraph_hashing_protocol_challenge.tex
\subsection{Challenge~(\ref{ichal:second_challenge}): Using $k$-wise Independence to Compute the Induced Subgraph $G[V_i]$ in \MPC}
\label{sec:computing_the_induced_subgraph}
For each sub-sampled set of vertices $V_i$, we need to compute $G[V_i]$,
i.e. we need to send all the edges in the induced subgraph $G[V_i]$ to the machine $M_i$.
Let $Q_u$ denote the set of all machines containing $u$.
Each edge $(u,w)$ then needs to be sent to all machines that contain both $u$ and $w$, $Q_u\cap Q_w$.
Naively, one could try to send the sets $Q_u$ and $Q_w$ to the edge $e=(u,w)$, for all $e\in E$.
However, this strategy could result in $Q_v$ being replicated $d(v)$ times. %
Since the expected size of $Q_v$ is $|Q_v| = \pV \mM$ %
the total expected memory usage of this strategy would be
$\sum_{v\in V} |Q_v|\cdot d(v) = \widetilde\Theta_{\eps}\left(m\cdot \pV \mM\right) = \widetilde{\omega}_{\eps}(m)$,
since $\pV = \widetilde{\Theta}(1/\sqrt \mM)$.
This defies our goal of optimal total memory.

Instead, we address this challenge by using globally known hash functions to sample the vertices on each machine.
That is, we let $h : V\times [\mM] \rightarrow \{ 0,1\}$  (formally presented in \cref{def:hash_function}) be a hash function known globally to all the machines.
Then we can compute the induced subgraphs $G[V_i]$  as follows.
\begin{tboxalg}{{\bf Compute-Induced-Subgraphs}}\label{alg:find_induced_subgraphs}
\begin{algorithmic}[1]
        \State $Q_v \gets \left\{ i\in [\mM] \mid h(v, i) = 1\right\}.$
        \State $Q_w \gets \left\{ i\in [\mM] \mid h(w, i) = 1\right\}.$
    \ParFor{$i \in Q_v\cap Q_w$}
        \State Send $e$ to machine $M_i$, containing $V_i$.
    \EndParFor
\end{algorithmic}
\end{tboxalg}
\begin{definition}
	\label{def:hash_function}
	The hash function $h(v, i)$ indicates whether vertex $v$ is sampled in $V_i$ or not.
	Specifically, $h : V\times [\mM] \rightarrow \{ 0,1\}$ such that $\mathbb{P}[h(v, i) = 1] = \pV$ for all $v\in V$ and $i\in [\mM]$.
	Recall that $\mM$ is the number of machines, and $\pV=\frac{1}{10}\cdot\sqrt{\frac{S}{mk}}$ is the sampling probability set in \cref{eq:define-pV}.
\end{definition}

\myparagraph{Using limited independence}
\label{par:using_limited_independence}
Ideally, we would want a perfect hash function, which would allow us to sample the $V_i$'s i.i.d.~from the uniform distribution on $V$.
However, since the hash function needs to be known globally, it must fit into each of the machines.
This implies that we \emph{cannot} use a fully independent perfect hash function.
Rather, we \emph{can} use one that has a high level of independence.
Specifically, given that the space per machine is $S$, we can have a globally known hash function $h$ that is $k$-wise independent\footnote{
A $k$-wise independent hash function is one where the hashes of \emph{any} $k$
distinct keys are guaranteed to be independent random variables
(see \cite{limited_independence_hash_function}).} for any $k < \Theta(S/\log n)$.
In fact, we can get away with as little as $(6\log n)$-wise independence (i.e., $k = 6\log n$).
Recalling \cref{eq:define-pV}, this also fixes the sampling probability to be $\pV = \sqrt{S/600m\log n}$.

%% file: 660-concentration-of-counts.tex
\subsubsection{Showing Concentration for the Triangle Count}
\label{sec:showing_concentration_for_the_triangle_count}

In the subsequent proofs, we will use the following assumptions from within \cref{thm:approximate-counting} (note that we added specific constants).
\begin{align}
\label{eq:estimator_thm_constraints}
T \ge 10\sqrt{\frac{mk}{S}}
&&S \ge \max\left\{15\frac{\sqrt{mk}}{\eps}, \frac{100kn^2}{m} \right\}
&&\mM = \frac{2000 mk}{\eps^2S}
\end{align}

Note that we set the number of machines to a specific value, instead of lower bounding it.
This is acceptable, because we can just ignore some of the machines.

\cref{alg:approx-count-triangles} outputs an estimate on the number of triangles in $G$ (\cref{line:approx-count-return}).
It is not hard to show that in expectation this output equals $T$ \emph{even with limited independence} as
discussed above.
The main challenge is to show that this output also concentrates well around its expectation.
Specifically, we show the following claim.
\begin{lemma}\label{lem:output-approx-count-Chebyshev}
Ignore \cref{line:check-size-of-G[V_i]} of \cref{alg:approx-count-triangles}.
Let $\hat T$ be as defined on \cref{line:approx-count-define-X} and $\mM = \frac{20}{\eps^2 \pV^2}$
be as defined in \cref{eq:estimator_thm_constraints}, and assume that $T \ge 1/\pV$.
Then, the following hold:
    \begin{enumerate}[(A)]
        \item\label{item:ee{X}} $\ee{\hat T} = \pV^3 \cdot R \cdot T$, and
        \item\label{item:X-ee{X}} $\pp{|\hat T - \ee{\hat T}| > \eps \ee{\hat T}} < \frac{1}{10}$.
    \end{enumerate}
\end{lemma}

\label{sec:proof-approx-count-Chebyshev}
We will prove Property~\eqref{item:X-ee{X}} of the claim by applying Chebyshev's inequality, for which we need to compute $\var{\hat T}$.
Let $\triangles{G}$ be the set of all triangles in $G$.
For a triangle $t \in \triangles{G}$, let $\hat T_{i, t} = 1$ if $t \in V[G_i]$, and $\hat T_{i, t} = 0$ otherwise.
Hence, $\hat T_i = \sum_{t \in \triangles{G}} \hat T_{i, t}$.
We begin by deriving $\ee{\hat T}$ and then proceed to showing that $\var{\hat T} = \sum_{i = 1}^R \var{\hat T_i}$.
After that we upper-bound $\var{\hat T_i}$ and conclude the proof by applying Chebyshev's inequality.

\paragraph{Deriving $\ee{\hat T}$.}
    Let $t$ be a triangle in $G$. Let $\hat T_t$ be a random variable denoting the total number of times $t$ appears in $G[V_i]$, for all $i = 1 \ldots R$. Given that $\pp{u \in V_i} = \pV$, we have that $\pp{t \in G[V_i]} = \pV^3$. Therefore, $\ee{\hat T_t} = R \cdot \pV^3$.

    Since $\hat T = \sum_{t \in \triangles{G}} \hat T_t$, we have
    \begin{equation}\label{eq:expectation-of-approx-count}
        \ee{\hat T} = \sum_{t \in \triangles{G}} \ee{\hat T_t} = \pV^3 \cdot R \cdot T.
    \end{equation}
    This proves Property~\eqref{item:ee{X}} of this claim.

    \paragraph{Decoupling $\var{\hat T}$.}
    To compute variance, one considers the second moment of a given random variable. So, to compute $\var{\hat T}$, we will consider products $\hat T_{i, t_1} \cdot \hat T_{j, t_2}$. Each of those products depend on at most $6$ vertices. Now, given that we used a $6$-wise independent function (see \cref{sec:computing_the_induced_subgraph}) to sample vertices in each $V_i$, one could expect that $\var{\hat T_i}$ and $\var{\hat T_j}$ for $i \neq j$ behave like they are independent, i.e., one could expect that it holds $\var{\hat T} = \sum_{i = 1}^R \var{\hat T_i}$. As we show next, it is indeed the case. We have
    \begin{eqnarray}
        \var{\hat T} & = & \ee{\hat T^2} - \ee{\hat T}^2 \nonumber \\
        & = & \ee{\rb{\sum_{i = 1}^R \sum_{t \in \triangles{G}} \hat T_{i, t}}^2} - \rb{\sum_{i = 1}^R \sum_{t \in \triangles{G}} \ee{\hat T_{i, t}}}^2 \label{eq:var-X-summands}
    \end{eqnarray}
    Consider now $\hat T_{i, t_1}$ and $\hat T_{j, t_2}$ for $i \neq j$ and some $t_1, t_2 \in \triangles{G}$ not necessarily distinct. In the first summand of \eqref{eq:var-X-summands}, we will have $\ee{2 \hat T_{i, t_1} \cdot \hat T_{j, t_2}}$. The vertices constituting $t_1$ and $t_2$ are $6$ \emph{distinct} copies of some (not necessarily all distinct) vertices of $V$. Since they are chosen by applying a $6$-wise independent function, we have $\ee{2 \hat T_{i, t_1} \cdot \hat T_{j, t_2}} = 2 \ee{\hat T_{i, t_1}} \cdot \ee{\hat T_{j, t_2}}$.\\
    On the other hand, the second summand of \eqref{eq:var-X-summands} also contains $2 \ee{\hat T_{i, t_1}} \cdot \ee{\hat T_{j, t_2}}$, which follows by direct expansion of the sum. Therefore, all the terms $\ee{2 \hat T_{i, t_1} \cdot \hat T_{j, t_2}}$ in $\var{\hat T}$ for $i \neq j$ cancel each other. So, we can also write $\var{\hat T}$ as
    \begin{eqnarray}
        \var{\hat T} & = & \sum_{i = 1}^R \ee{\rb{\sum_{t \in \triangles{G}} \hat T_{i, t}}^2} - \sum_{i = 1}^R \rb{\sum_{t \in \triangles{G}} \ee{\hat T_{i, t}}}^2 \nonumber \\
        & = & \sum_{i = 1}^R \var{\hat T_i}. \label{eq:var-X-sum-of-vars}
    \end{eqnarray}
    Therefore, to upper-bound $\var{\hat T}$ it suffices to upper-bound $\var{\hat T_i}$.

    \paragraph{Upper-bounding $\var{\hat T_i}$.}
        We have
        \begin{eqnarray}
            \var{\hat T_i} & = & \ee{\rb{\sum_{t \in \triangles{G}} \hat T_{i, t}}^2} - \rb{\sum_{t \in \triangles{G}} \ee{\hat T_{i, t}}}^2 \nonumber \\
            & \le & \ee{\rb{\sum_{t \in \triangles{G}} \hat T_{i, t}}^2} \nonumber  \\
            & = & \ee{\sum_{t \in \triangles{G}} \hat T_{i, t}^2} + \ee{\sum_{t_1, t_2 \in \triangles{G}; t_1 \neq t_2} \hat T_{i, t_1} \cdot \hat T_{i, t_2}}. \label{eq:var-Xi-two-summands}
        \end{eqnarray}
        Since each $\hat T_{i, t}$ is a $0/1$ random variables, $\hat T_{i, t}^2
        = \hat T_{i, t}$. Let $t_1 \neq t_2$ be two triangles in
        $\triangles{G}$. Let $k$ be the number of distinct vertices they are
        consisted of, which implies $4 \le k \le 6$. Then, observe that
        $\ee{\hat T_{i, t_1} \cdot \hat T_{i, t_2}} = \pV^k \le \pV^4$. We now have all ingredients to upper-bound $\var{\hat T_i}$. From \eqref{eq:var-Xi-two-summands} and our discussion it follows
        \begin{equation}\label{eq:var-Xi-final-bound}
            \var{\hat T_i} \le T \pV^3 + T^2 \pV^4 \le 2 T^2 \pV^4,
        \end{equation}
        where we used our assumption that $T \ge 1/\pV$.

    \paragraph{Finalizing the proof.}
        From \eqref{eq:var-X-sum-of-vars} and \eqref{eq:var-Xi-final-bound} we have
        \[
            \var{\hat T} \le 2 R T^2 \pV^4.
        \]
        So, from Chebyshev's inequality and \eqref{eq:expectation-of-approx-count} we derive
        \begin{eqnarray*}
            \pp{|\hat T - \ee{\hat T}| > \eps \ee{\hat T}} & < & \frac{\var{\hat T}}{\eps^2 \ee{\hat T}^2} \\
            & \le & \frac{2 R T^2 \pV^4}{\eps^2 \pV^6 R^2 T^2} \\
            & = & \frac{2}{\eps^2 \pV^2 R}.
        \end{eqnarray*}
        Hence, for $R \ge \frac{20}{\eps^2 \pV^2}$ we get the desired bound.

%% file: 320-protocol_sub_challenge.tex
\subsection{Challenge~(\ref{ichal:bound_msgs_challenge}): Showing that, with high constant probability, the size of the sent/received messages is bounded.}
\label{sec:sub_challenge_showing_that_protocol_succeeds}

We need to show that the number of edges sent and received by any machine $M_i$ is at most $S$ \emph{with high constant probability}.
To this end, we partition the vertex set $V$ into $V_{light}$ and $V_{heavy}$ by picking a threshold degree $\tau$ for the vertices.
Following this, we define \emph{light edges} as ones that have both end-points in $V_{light}$,
and conversely, any edge with at least one end-point in $V_{heavy}$ is designated as \emph{heavy}.
In order for the protocol to suceed, the following must hold:
\begin{enumerate}[(A)]
    \item The number of \emph{light edges} concentrates (see \cref{sec:light_vertices}).
    \item The number of \emph{heavy edges} concentrates (see \cref{sec:heavy_vertices}).
    \item The number of sent messages is at most $S$ (see \cref{sec:bounding_the_number_of_messages_sent_by_any_machine}).
\end{enumerate}

The first two items ensure that each machine $M_i$ \emph{receives} at most $S$ messages,
and the last item ensures that each machine \emph{sends} at most $S$ messages.
Given the above, we proceed to address the last challenge.

%% file: 620-bounding_light_edges.tex
\subsubsection{Bounding the Number of Light Edges \emph{Received} by a Machine}
\label{sec:bounding_the_number_of_light_edges_on_a_machine}
We will now bound the probability that any of the induced subgraphs \emph{does not fit} on a machine.
To that end, we set a degree threshold $\tau = \frac{k}{\pV}$,
and define the set of \emph{light} vertices $V_{light}$ to be the ones with degree less than $\tau$.
All other vertices are \emph{heavy}, and we let them comprise the set $V_{heavy}$.

Fix a machine $M_i$.
We prove that, with probability at least $9/10$, the number of edges in $G[V_i]$ is upper bounded by $S$.

We start with analyzing the contribution of the light vertices to  the induced subgraphs.
We first consider the simpler case of bounding the number of edges in $G[V_i]$ that have both end-points in $V_{light}$. We refer to such edges as \emph{light edges} and denote them by $E_{light}.$
 For every edge $e\in E_{light}$, we define a random variable $Z^{(i)}_e$ as follows.
\label{sec:light_vertices}
$$
Z^{(i)}_e =
\begin{cases}
1 &\textrm{if } e\in G[V_i],\\
0 &\textrm{otherwise}.
\end{cases}
$$

We let $Z^{(i)}$ be the sum over all  random variables $Z^i_e$, $Z^i=\sum_{e \in E_{light}}Z^i_e$, and we let  $m_{\ell}$ denote the total number of edges with \emph{light} endpoints in the original graph $G$, i.e., $m_{\ell} = |E_{light}|$.

We prove the following lemma.
\begin{lemma}\label{lem:bound-Z}
	With probability at least $9/10,$ for every $i\in [\mM]$, $G[V_i]$ contains at most $\frac{1}{4}S$ light edges.
\end{lemma}
\begin{proof}
Fix a machine $M_i$, and let $Z=Z^i$ be as defined in the previous paragraph.
\begin{align*}
\mathbb E[Z] &= \mathbb E\left[ \sum\limits_{e\in E_{light}} Z_e\right] = m_{\ell}\hat{p}^2
\le m\cdot \frac{S}{100mk} = \frac{S}{100k} \le \frac{S}{100}\;.
\end{align*}
As $Z_e$ are $\{0,1\}$ random variables, we also have $\ee{Z} = \ee{\sum\limits_{e\in E_{light}} Z_e^2}$. Now we upper-bound the variance.
\begin{align*}
\var{Z} &= \mathbb E\left[ \left(\sum\limits_{e\in E_{light}} Z_e\right)^2\right] - \mathbb E\left[ \sum\limits_{e\in E_{light}} Z_e\right]^2 \\
&\le \sum\limits_{e\in E_{light}} \mathbb E[Z_e^2]
+ \sum\limits_{\substack{e_1, e_2\in E_{light}\\ e_1 \not= e_2}} 2\cdot \mathbb E\left[ Z_{e_1}Z_{e_2}\right] \\
& \quad - \sum\limits_{\substack{e_1, e_2\in E_{light}\\ e_1 \not= e_2}} 2\cdot \mathbb E[ Z_{e_1}]\mathbb E[Z_{e_2}] \\
&= m_{\ell}\cdot \hat{p}^2
+ \sum\limits_{\substack{e_1, e_2\in E_{light}\\ e_1 \not= e_2}} 2\cdot \mathbb E\left[ Z_{e_1}Z_{e_2}\right] \\
& \quad - \sum\limits_{\substack{e_1, e_2\in E_{light}\\ e_1 \not= e_2}} 2\cdot \mathbb E[ Z_{e_1}]\mathbb E[Z_{e_2}] \\
&\le m_{\ell}\cdot \hat{p}^2
+ \sum\limits_{\textrm{$e_1$ and $e_2$ intersect}} 2\cdot \mathbb E\left[ Z_{e_1}Z_{e_2}\right] \\
&\le m_{\ell}\cdot \hat{p}^2 + \left( \sum\limits_{v\in V_{light}} d(v)^2\right)\cdot \hat{p}^3 \\
&\le m_{\ell}\cdot \hat{p}^2 + \left( \sum\limits_{v\in V_{light}} d(v)\right)\cdot \frac{k}{\hat{p}}\cdot \hat{p}^3 \\
&\le 3m_{\ell}\cdot \hat{p}^2\cdot k \le 3m\cdot \frac{S}{100mk}\cdot k < \frac{S}{30}
\end{align*}
\end{proof}

We can now use Chebyshev's inequality to conclude that
\begin{align*}
\pp{ |Z^{(i)}-\mathbb E[Z^{(i)}]| > S/\sqrt 3} &\le \frac{\var{Z^{(i)}}}{S^2/3} \\
\implies \pp{ Z^{(i)} > 3S/4} &\le \frac{3}{30S} = \frac 1{10S}
\end{align*}

Finally, we can use union bound over all $\mM$ machines to upper bound the probability that,
\emph{any} of the $Z^{(i)}$ values exceeds $3S/4$ (using the the constraints descrbed in \cref{eq:estimator_thm_constraints} to simplify).
\[
\frac{\mM}{10S}
= \frac{2000mk}{\eps^2 S}\cdot \frac{1}{10S}
\le \frac{200mk}{\eps^2}\cdot \frac{1}{(15\sqrt{mk}/\eps)^2}
= \frac{200 mk}{\eps^2 S^2},
\]
Therefore, with probability at least $9/10$, none of the induced subgraphs $G[V_i]$ will contain more than $3S/4$ light edges.

%% file: 630-bounding_heavy_edges.tex
\subsubsection{Bounding the Number of Heavy Edges \emph{Received} by a Machine}
\label{sec:bounding_the_number_of_heavy_edges_on_a_machine}
Next, we turn our attention to the edges that have at least one endpoint in $V_{heavy}$ (we call such edges \emph{heavy}).
We will show that for each $v\in V_{heavy} \cap  V_i$,
the number of edges contributed by $v$ concentrates around its expectation.\footnote{
Intuitively, this is because $v$ has high degree, and therefore the number of its sampled neighbors ($|N(v)\cap V_i|$) will concentrate.}
In this section, we will use $2m_{h}$ to denote the total degree of all the heavy vertices i.e. $2m_{h} = \sum_{v\in V_{heavy}}d(v)$.

Let $Z^{(v)}_w$ be the $\{ 0,1\}$ indicator random variable for $w\in V_i$ conditioned on the event that $v\in V_i\cap V_{heavy}$.
We use this conditioning on $v$ being present, because, in its absence, the number of edges contributed by $v$,
can be \emph{zero} with probability $(1-\pV)$, i.e. this naive estimator would not concentrate around its expectation.

Let $Z^{(v)}$ be the sum of all $Z^{(v)}_w$ for $w\in N(v)$.
For a particular $v$, the $Z^{(v)}_w$ variables are $k$-wise independent, which allows us to use the following lemma to bound $Z^{(v)}$.
In what follows, we will omit the super-script $(v)$ for the sake of convenience.
\label{sec:heavy_vertices}
\begin{lemma}
\label{lem:limited_independence_concentration}
If $Z_1, Z_2,\cdots, Z_n$ are $k$-wise independent $\{ 0,1\}$ random variables with $\mathbb E[Z_i] = p$ and $k \le np$, then for $Z = \sum_i Z_i$ we have
\[
\pp{Z > 3np} \le 2^{-k}.
\]
\end{lemma}
\begin{proof}
To prove the claim, we will re-write $\mathbb P[\sum Z_i > 3np]$, as the probability that the number of size $k$ subsets
of $\{ Z_1, Z_2,\cdots, Z_n\}$ that are all equal to $\mathsf 1$ is larger than $\binom{3np}{k}$.
\begin{align*}
& \pp{Z > 3np}\\
&= \pp{\left|\{ T\ :\ T \subseteq [n], |T| = k, \text{ and }Z_i = 1 \ \forall i\in T\}\right| > \binom{3np}{k}} \\
&\le \frac{\ee{|\{T\ :\ T\subseteq [n], |T| = k, \text{ and } Z_i = 1 \ \forall i\in T\}|}}{\binom{3np}{k}} \\
& = \frac{\binom{n}{k}\cdot p^k}{\binom{3np}{k}} \le \left( \frac{n}{3np-k}\cdot p\right)^k \le \left( \frac{np}{2np}\right)^k = 2^{-k}
\end{align*}
where to obtain $3np - k \ge 2 np$ we used our assumption that $k \le np$.
\end{proof}

Since $v$ is heavy, there are at least $\tau$ variables in the sum $Z^{(v)} = \sum_{w\in N(v)}Z^{(v)}_w$.
Additionally, we know that $\mathbb E[Z^{(v)}_w] = \pV$ and $k\le \tau\pV$.
Thus, we obtain the following corollary from \cref{lem:limited_independence_concentration}:
\begin{corollary}
\label{cor:high_degree_vertex_edge_concentration}
For any vertex $v\in V_{heavy}\cap V_i$, we get $\pp{Z^{(v)} > 3d(v)\cdot \hat p} < 2^{-k}$, or explicitly
\[
\pp{N(v) \cap V_i > 3d(v)\hat p \mid v\in V_i \text{ and } d(v) > \tau} < 2^{-k} = \frac{1}{n^6}
\]
\end{corollary}
\begin{corollary}
\label{cor:all_high_degree_vertices_edge_concentration}
With high probability $1-\frac1{n^{5}}$, we ensure that for all $v\in V_{heavy}$, $Z^{(v)} \le 3\cdot \mathbb E\left[Z^{(v)}\right]$
\end{corollary}

The important point is that the sum of $Z^{(v)}$ (over all $v\in V_i$) is an upper bound on $m_{h}$ -- the number of heavy edges in $G[V_i]$.
In order to bound this sum, we define random variables $W_v$ for each $v\in V_{heavy}$ as follows:
$$
W_v =
\begin{cases}
\frac{d(v)}{n} &\textrm{if } v\in V_i\\
0 &\textrm{otherwise}
\end{cases}
$$

We also define $W$ to be the sum of all $W_v$,
thus implying $\mu = \mathbb E[W] = \sum\limits_{v\in V_{heavy}}\pV\cdot\frac{d(v)}{n}\le\frac{2\pV m_{h}}{n}$.
\begin{theorem}
\label{thm:limited_independence_concentration_continuous}
(Theorem 5 from \cite{limited_independence})
If $W$ is the sum of $k$-wise independent random variables, each of which takes values in the interval $[0,1]$,
and $\delta \ge 1$, then:
\begin{align*}
k < \floor{\delta\mu e^{-1/3}} &\implies \pp{ |W-\mu| > \delta\mu} \le e^{\floor{k/2}}
\end{align*}
\end{theorem}

\begin{corollary}
\label{cor:concentration_of_W}
$\pp{ W > \frac{4\hat{p}mk}{n} } \le e^{-\floor{k/2}}$
\end{corollary}
\begin{proof}
We can use the fact the random variables $W_v$ are $k$-wise independent to apply \cref{thm:limited_independence_concentration_continuous}.
First, we ensure that $k < \floor{\delta\mu e^{-1/3}}$, that we achieve by setting $\delta = \frac{mk}{m_{h}}$.

Recall that $m_{h}$ is the number of heavy edges (ones with at least one heavy end-point), and $m$ is the total number of edges in the original graph $G$.
\[
\delta = \frac{mk}{m_{h}} \implies \delta\mu e^{-1/3} = \frac{mk\cdot 2\hat{p}m_{h}}{m_{h}\cdot n}\cdot e^{-1/3} > \frac{\hat{p}mk}{n}
\implies \delta\mu e^{-1/3} > k
\]
In the last step, we used the fact that $S > 100kn^2/m$ from \cref{eq:estimator_thm_constraints}, to imply that $\pV m/n > 1$.
Therefore, we can now apply \cref{thm:limited_independence_concentration_continuous} to conclude:
\begin{align*}
\pp{ |W-\mu| > \delta\mu} &\le e^{-\floor{k/2}} \\
\implies \pp{ W > \mu + \frac{2\hat{p}mk}{n} } &\le e^{-\floor{k/2}} \\
\implies \pp{ W > \frac{4\hat{p}mk}{n} } &\le e^{-\floor{k/2}}
\end{align*}
In the second step, we used the fact that $\mu = \mathbb E[W] = \sum\limits_{v\in V_{heavy}} \hat p\cdot \frac{d(v)}{n} \le \frac{2m \hat p}{n}$.
\end{proof}

Now we are finally ready to upper bound the number of heavy edges in $G[V_i]$.
With high probability (using \cref{cor:high_degree_vertex_edge_concentration}), the following holds:
\begin{align*}
\#\left( \textrm{heavy edges in } G[V_i] \right) &\le \mathlarger\sum\limits_{v\in V_{heavy}} \pp{v\in V_i}\cdot (3d(v) \hat{p}) \\
&\le \mathlarger\sum\limits_{v\in V_{heavy}} W_v\cdot n \cdot (3 \hat{p}) = 3n\hat{p}\cdot W \\
&\le 12 \hat p^2 mk = \frac{12S}{100} < \frac S8
\end{align*}

\begin{theorem}[Heavy edges]
\label{thm:heavy_edge_concentration}
With high probability, the number of edges in $G[V_i]$ that have some endpoint with degree larger than $\tau$ is at most $S/8$.
\end{theorem}

Combining this result with \cref{thm:heavy_edge_concentration}, we conclude the following:
\begin{theorem}
\label{thm:induced_subgraph_edge_concentration}
With probability at least $9/10$, the maximum number of edges in any of the $G[V_i]$s (where $i\in [R]$) does not exceed $S$,
and hence \cref{alg:approx-count-triangles} does not terminate on \cref{line:check-size-of-G[V_i]}.
\end{theorem}

%% file: 640-sent_messages.tex
\subsubsection{Upper-Bounding the Number of Messages \emph{Sent} by any Machine}
\label{sec:bounding_the_number_of_messages_sent_by_any_machine}

Recalling \cref{alg:find_induced_subgraphs}, we note that the number of messages \emph{received} by the machine containing $V_i$,
is equal to the number of edges in $G[V_i]$.
Therefore, the last section essentially proved that the number of messages (edges) \emph{received} by a particular machine is upper-bounded by $S$.
Conversely, in this section, we will justify that the number of messages \emph{sent} by any machine is $O(S)$.
Since the number of edges stored in a machine is $\le S$, it suffices to to show that for each edge $e$,
\cref{alg:find_induced_subgraphs} sends only $O(1)$ messages (each message is a copy of the edge $e$).

Let $Z^{(e)}_i$ be the $\{ 0,1\}$ indicator random variable for $e\in G[V_i]$, and let $Z^{(e)}$ be the sum of $Z^{(e)}_i$ for all $i\in [\mM]$.
Here, $Z^{(e)}$ represents the number of messages that are created by edge $e$.
Additionally we make $r = S\mM/m = O_{\eps}(\log n)$ copies of each edge $e$, and ensure that all replicates reside on the same machine.
We distribute the $Z^{(e)}$ messages evenly amongst the replicates, so that each replica is only responsible for $Z^{(e)}/r$ messages.

Since all replicates are on the same machine, this last step is purely conceptual,
but it will simplify our arguemnt, by allowing us to charge the outgoing messages to each replicate (as opposed to each edge).
Our goal will be show that each replicate is responsible for only $O(1)$ messages,
which is the same as showing that w.h.p. $Z^{(e)}/r = O(1)$.

Clearly $\mu = \mathbb E[Z^{(e)}] = \pV^2\cdot \mM = \frac{S\mM}{100mk}$.
This allows us to apply \cref{lem:limited_independence_concentration} with $\delta = \frac{100e^{1/3}mk^2}{S\mM}$
\begin{align*}
\mathbb P\left[ Z^{(e)} > \delta\mu\right] \le e^{-\floor{k/2}} = \frac{1}{n^3}
\implies \mathbb P\left[ \frac{Z^{(e)}}{r} > \frac{e^{1/3}k}{r} \right] \le \frac{1}{n^3}
\end{align*}
Using the assumption (from \cref{eq:estimator_thm_constraints}) that $\mM > 2000mk/S\implies r > 2000k$,
we see that with high probability, the number of messages sent by any replicate is bounded above by $e^{1/3}/2000\le 1$.
So, the number of messages sent from any machine is bounded by $S$ with high probability.

%% file: 330-num_triangles_challenge.tex
\subsection{Challenge~(\ref{ichal:third_challenge}): $\hat T$ is close to its expected value}
\label{sec:estimating_number_of_triangles_from_the_aggregate}

In this section, we provide merely a brief discussion of this challenge for intuition,
and we fully analyze the approximation guarantees of our algorithm in \cref{sec:showing_concentration_for_the_triangle_count}.
That analysis also makes clear the source of our advertised lower-bound on $T$ for which an estimated count concentrates well.

\myparagraph{Lower Bound on Number of Triangles}
\label{par:lower_bound_on_number_of_triangles}
In order to output \emph{any} approximation
(note that we are ignoring all factors of $\eps$ and $O(\poly \log n)$ here)
to the triangle count,
we must see $\Omega(1)$ triangles amongst all of the induced subgraphs on all the machines.
The expected number of triangles in a specific induced $G[V_i]$ is $\pV^3 T$,
and therefore, the expected number of triangles overall is $\pV^3 T\mM$
which must be $\Omega(1)$ for some setting of $T$.
Since we set $\pV$ such that $\pV^2 m = \Theta(S)$, this gives
that $\pV^2 = O(S/m)$ which implies
$\pV^2\cdot \mM = \pV^2 \cdot (m/S) = \Theta(1)$.
This then immmediately implies that to show that $\pV^3 T$ is
$\Omega(1)$, we need only show that $\pV\cdot T$ is $\Omega(1)$.
Specifically, we show in~\cref{lem:output-approx-count-Chebyshev} that when $T > 1/\pV$,
we can obtain a $(1\pm\eps)$-approximation.
To get some intuition for this lower bound on $T$, note that, in the linear memory regime, when $S = \Theta(n)$,
this translates to $T>\sqrt{d_{avg}}$, where $d_{avg}$ is the average degree of $G$.
\[
T > \frac{1}{\pV} = \widetilde\Theta\left(\sqrt{\frac{m}{S}}\right)
\xRightarrow{\quad\textrm{for } S = \widetilde\Theta(n)\quad} T > \widetilde\Theta\left(\sqrt{d_{avg}}\right).
\]

%% file: 680-high_probability_bound.tex
\subsubsection{Getting the High Probability Bound}
\label{sec:getting_the_high_probability_bound}

By building on \cref{lem:output-approx-count-Chebyshev} and \cref{alg:approx-count-triangles}, we design \cref{alg:approx-count-triangles-main} that outputs an approximate triangle counting with high probability, as opposed with only constant success probability. It is important to note that 
in the below algorithm, all $O(\log n)$ \emph{independent iterations}
(\cref{line:parallel}) are done \emph{in parallel}, simultaneously, not sequentially.

\begin{tboxalg}{{\bf Approximate Triangle Counting}}\label{alg:approx-count-triangles-main}
\begin{algorithmic}[1]
  \Function{Approx-Triangles-Main}{$G = (V, E)$}

    \State Let $I \gets 100 \cdot \log{n}$. \label{line:alg-approx-count-define-I}

    \ParFor{$i \gets 1 \ldots I$}\Comment{Perform all $I$ iterations in \emph{parallel} simultaneously in $O(1)$ rounds.}\label{line:parallel}
        \State Let $Y_i$ be the output of \cref{alg:approx-count-triangles} invoked on $G$. We assume that each invocation of \cref{alg:approx-count-triangles} uses fresh randomness compared to previous runs.\label{line:alg-approx-count-define-Yi}
    \EndParFor

    \State Let $\cY$ be the list of all $Y_i$, for $i = 1 \ldots I$.
    \State Sort $\cY$ in non-decreasing order.
    \State \Return the median of $\cY$
\EndFunction
\end{algorithmic}
\end{tboxalg}
We have the following guarantee for \cref{alg:approx-count-triangles-main}.
\begin{theorem}
\label{thm:median_trick}
    Let $Y$ be the output of \cref{alg:approx-count-triangles-main}. Then, with high probability it holds
    \[
        |Y - T| \le \eps T.
    \]
\end{theorem}
In the proof of this theorem we use the following concentration bound.
\begin{theorem}[Chernoff bound]\label{lem:chernoff}
	Let $X_1, \ldots, X_k$ be independent random variables taking values in $[0, 1]$. Let $X \eqdef \sum_{i = 1}^k X_i$ and $\mu \eqdef \ee{X}$. Then, or any $\delta \in [0, 1]$ it holds $\pp{X \le (1 - \delta) \mu} \le \exp\rb{- \delta^2 \mu / 2}$.
\end{theorem}
\begin{proof}[Proof of \cref{thm:median_trick}]
    The proof of this theorem is essentially the so-called ``Median trick''. We provide full proof here for completeness.

    Let $Y_i$ be as defined on \cref{line:alg-approx-count-define-Yi} of \cref{alg:approx-count-triangles-main}. By \cref{thm:induced_subgraph_edge_concentration},
    with probability at most $1/10$ \cref{alg:approx-count-triangles} terminates due to creating too big subgraphs. If we ignore \cref{line:check-size-of-G[V_i]} of \cref{alg:approx-count-triangles}, then by Property~\eqref{item:ee{X}} of \cref{lem:output-approx-count-Chebyshev} we have $\ee{Y_i} = T$. $Y_i$ significantly deviates from its expectation if \cref{alg:approx-count-triangles} terminates on \cref{line:check-size-of-G[V_i]} or if the  estimate $Y_i$ is simply off. Define a $0/1$ variable $Z_i$ which equals $1$ iff $|Y_i - T| \le \eps T$. By union bound on Property~\eqref{item:X-ee{X}} of  \cref{lem:output-approx-count-Chebyshev} and \cref{thm:induced_subgraph_edge_concentration}, we have $\pp{Z_i = 1} \ge 1 - 1/10 - 1/10 = 4/5$. Also, following \cref{line:alg-approx-count-define-Yi} of \cref{alg:approx-count-triangles-main} we have that all $Z_i$ are independent.

    Let $Z = \sum_{i = 1}^I Z_i$. We have that $\ee{Z} \ge \tfrac{4}{5} I$, implying that in expectation at least $4/5$ fraction on $Z$-variables are $1$. We now bound the probability that at least $2/5$ of these variables equal $0$, i.e, at most $3/5$ of them equal $1$. Since $Z$-variables are independent, for this we can use \cref{lem:chernoff}, obtaining
    \[
        \pp{Z \le \frac{3}{5} I} \le \pp{Z \le \rb{1 - \frac{1}{5}} \ee{Z}} \le \exp{\rb{-\ee{Z} / 50}}.
    \]
    Given that $I = 100 \cdot \log{n}$ (see \cref{line:alg-approx-count-define-I} of \cref{alg:approx-count-triangles-main}), we derive that $\pp{Z \le \frac{3}{5} I} < n^{-1}$. This now implies that with probability at least $n^{-1}$ the output of \cref{alg:approx-count-triangles-main} is some $Y_j$ such that $Z_j = 1$. This completes the analysis.
\end{proof}

%% file: k-subgraph.tex
\subsection{Showing Concentration for the $\sub$-Subgraph Count}\label{sec:k-subgraph}

Using similar analysis to the previous section, in this section,
we show the expectation and concentration bound of our
subgraph counting algorithm for any subgraphs consisting of $\sub$ nodes where
$\sub$ is constant. Let this subgraph be $H$.

\begin{lemma}\label{lem:expected-subgraphs}
Let $\hat \countest$ be the count of subgraph $H$ (with $\sub$ vertices)
in $G[V_i]$
and $\mM = \frac{20}{\eps^2 \pV^{K - 1}}$
be as defined in \cref{eq:estimator_thm_constraints}, and assume that
$\countest \ge 1/\pV$.
Then, the following hold:
    \begin{enumerate}[(A)]
        \item\label{item:k-A} $\ee{\hat \countest} = \pV^{\sub} \cdot R \cdot
        \countest$, and
        \item\label{item:k-B} $\pp{|\hat \countest - \ee{\hat \countest}| >
        \eps \ee{\hat \countest}} < \frac{1}{10}$.
    \end{enumerate}
\end{lemma}

\begin{proof}
We first prove~\cref{item:k-A}. The probability that a particular
$\sub$-vertex
occurrence $h$ of $H$ appears in machine
$M_i$ is $\pp{h \in G[V_i]} = \pV^{\sub}$.
There are $\countest$ number of occurrences of $H$ in $G$. Thus,
    the expected number of occurrences of $H$ in machine $M_i$ is $\ee{\hat
    \countest_{M_i}} = \pV^{\sub} \cdot
\countest$. Since there are $R$ machines (which did not exceed the memory limit),
the expected number of occurrences of $H$ in all $R$ machines is
$\ee{\hat \countest} = \sum_{i \leq
R} \ee{\hat \countest_{M_i}} = \pV^{\sub} R
\countest$.

We now prove~\cref{item:k-B}. Let $H(G)$ be the set of occurrences
of $H$ in $G$. Let $\hat \countest_{i, h}$ be a random variable where $\hat
\countest_{i, h} = 1$ if $h$, a particular occurrence of $H$ in $G$, is in
machine $i$; otherwise, $\hat \countest_{i, h} = 0$.
Then,

\begin{eqnarray}
    \var{\hat \countest} &=& \ee{\hat \countest^2} - \ee{\hat \countest}^2\\
    &=& \ee{\left(\sum_{i = 1}^R \sum_{h \in H(G)}\hat \countest_{i, h}\right)^2}
    - \left(\sum_{i = 1}^R \sum_{h \in H(G)} \ee{\hat \countest_{i,
    h}}\right)^2.\label{eq:var-k}
\end{eqnarray}

First, consider random variables $\hat \countest_{i, h_1}$ and $\hat
\countest_{j, h_2}$; for each $i \neq j \in [R]$ and each $h_1, h_2 \in H(G)$, there
exists a term in the first summand of~\cref{eq:var-k} containing
$\ee{2\hat{\countest}_{i, h_1} \hat{\countest}_{j, h_2}}$. The vertices
constituting $h_1$ and $h_2$ are $2\sub$ distinct copies of some not
necessarily distinct copies of vertices in $V$. Suppose we use a $2k$-wise
independent hash function, then we have $\ee{2\hat{\countest}_{i, h_1} \cdot
\hat{\countest}_{j, h_2}} = 2\ee{\hat{\countest}_{i, h_1}} \cdot
\ee{\hat{\countest}_{j, h_2}}$. We see that this term also shows up in the
second summand of~\cref{eq:var-k}. Hence, the terms cancel for each $i \neq j$ and we can
simplify~\cref{eq:var-k} to the following.

\begin{eqnarray}
\var{\hat{\countest}} = \sum_{i = 1}^R \ee{\left(\sum_{h \in H(G)}
\hat{\countest}_{i, h}\right)^2} - \sum_{i = 1}^R \left(\sum_{h \in H(G)}
\ee{\hat{\countest}_{i, h}}\right)^2 = \sum_{i = 1}^R
\var{\hat{\countest}_i}.\label{eq:k-var}
\end{eqnarray}

Now, what remains is to upper bound $\var{\hat{\countest}_i}$. Using the same
approach as in the previous section with the observation that any two distince
occurrences $h_1$ and $h_2$ must contain $\sub + 1 \leq k \leq 2\sub$ distinct
vertices. This means that $\ee{\hat{\countest}_{i, h_1} \cdot
\hat{\countest}_{i, h_2}} = \pV^{k} \leq \pV^{\sub + 1}$.
Then, we can bound
\begin{eqnarray}
\var{\hat{\countest}_i} \leq 2\countest^2 \pV^{\sub+
1}\label{eq:individual-k-var}
\end{eqnarray}
(assuming $\countest \geq 1/\pV$).

Then, from~\cref{eq:k-var} and~\cref{eq:individual-k-var}, we get the bound on
the variance to be $\var{\hat{\countest}} \leq 2R\countest^2 \pV^{\sub+1}$.

By Chebyshev's inequality and~\cref{item:k-A}, we compute
\begin{eqnarray}
\pp{|\hat{\countest} - \ee{\hat{\countest}}| > \eps \ee{\hat{\countest}}} <
\frac{\var{\hat{\countest}}}{\eps^2 \ee{\hat{\countest}}^2} \leq
\frac{2\countest^2 \pV^{\sub + 1}}{\eps^2 \pV^{2\sub} R^2
\countest^2} = \frac{2}{\eps^2 \pV^{K - 1} R^2}.
\end{eqnarray}

When setting $R \geq \frac{20}{\eps^2 \pV^{K - 1}}$, we obtain~\cref{item:k-B}.
\end{proof}

Using~\cref{lem:expected-subgraphs}, we can obtain the following theorem about
counting occurrences of \emph{any} subgraph $H$ with $\sub$ vertices.

\begin{theorem}\label{thm:subgraph}
	Let $G=(V, E)$ be a graph over $n$ vertices, $m$ edges, and let $\countest$
    be the number of occurrrences of subgraph $H$ with $\sub$ vertices
    in $G$.
	Assuming
	\begin{multicols}{2}
		\begin{enumerate}[(i)]
            \item $\countest = \widetilde\Omega\left(
                \left(\frac{m}{S}\right)^{\sub/2 - 1}\right)$,
			\item $S = \widetilde{\Omega}\left( \max\left\{\frac{\sqrt{m}}{\eps}, \frac{n^2}{m} \right\} \right)$,
		\end{enumerate}
	\end{multicols}
	there exists an $\MPC$ algorithm, using $\mM$ machines, each with local space $S$, and total space $\mM S = \tilde O_{\eps}(m)$,
	that outputs a $(1\pm \eps)$-approximation of $B$, with high probability, in $O(1)$ rounds.
\end{theorem}

\begin{corollary}\label{cor:k-linear-space}
    Let $G = (V, E)$ be an input graph and $\countest$ be the number of
    occurrences of subgraph $H$ with $\sub$ vertices in $G$. If $B \geq
    \davg^{\sub/2 - 1}$, then there exists an MPC algorithm that in $O(1)$
    rounds with high probability outputs a $(1+\eps)$-approximation of
    $\countest$. This algorithm uses a total space of $\tilde{O}(m)$ and space
    $\tilde{\Theta}(n)$ per machine. $\davg$ is the average degree of the
    vertices in the graph.
\end{corollary}

%% file: 400-five-subgraph.tex
\newcommand{\dG}{\overrightarrow{G}}
\newcommand{\dH}{\overrightarrow{H}}
\newcommand{\dE}{\overrightarrow{E}}
\newcommand{\mQ}{\mathcal{Q}}
\newcommand{\coreAlg}{\hyperref[core-alg]{\textup{\color{black}{\sf $sk$-Core-Orientation}}}}
\newcommand{\subCnt}{\hyperref[alg:sub-cnt]{\textup{\color{black}{\sf Sub-Cnt$_{\leq 5}$}}}}
\newcommand{\HM}{\mathcal{HM}}
\newcommand{\dT}{\overrightarrow{T}}
\newcommand{\dpp}{d^+}
\newcommand{\thh}{^{\textrm{th}}}

\section{Counting  subgraphs of size at most $5$ in bounded arboricity graphs}\label{sec:5-subgraphs}

In this section, we present a procedure that for every subgraph $H$ such that $|H|\leq 5$, counts the exact  number of occurrences of $H$ in $G$ in $O(\sqrt{\log n})$ rounds and $O(m\alpha^3)$ total memory, where as before, $\alpha$ is an upper bound on the arboricity of $G$ \footnote{Strictly speaking, we will have $\alpha\leq 5\alpha(G)$ but as this does not affect the asymptotic bounds, it is easier to just relate to it as the exact arboricity.}. The procedure is based on a recent paper by Bera, Pashanasangi and Seshadhri~\cite{BPS20} (henceforth BPS) which presented an $O(m\alpha^3)$ time and space algorithm for the same task in the sequential model.  We will start by a short description of the BPS result, and then continue to explain how to implement it in the MPC model.

\subsection{The BPS algorithm}
BPS generalize the ideas of Chiba and Nisheziki~\cite{CN85} for counting constant-size-cliques and 4-cycles in the classical sequential model to counting all subgraphs of up to $5$ nodes in $O(n+m\alpha^{3})$ time.
Let $H$ be the subgraph in question.
The main idea of BPS is as follows.  The algorithm starts by computing a degeneracy ordering of $G$,  which is an acyclic orientation of $G$, denoted $\dG$,  where each vertex has at most $O(\alpha)$ outgoing neighbors. The idea is then to consider all acyclic orientations of $H$ (up to isomorphisms), and for each such acyclic orientation $\dH$, count the number of occurrences of $\dH$ in $\dG$, as described next. The algorithm computes what is referred to as a largest directed rooted tree subgraph (DRTS) of $\dH$, denoted $\dT$. That is, the DRTS $\dT$ is a largest (in number of vertices) tree that is contained in $\dH$ such that all of the edges are directed away from the root of $\dT$. Given a DRTS $\dT$,  proceed by looking for all copies of $\dT$ in $\dG$. Once a copy of $\dT$ is found, it needs to be verified whether it can be extended to a copy of $\dH$ in $\dG$. This verification is based on the observation that for any directed subgraph $\dH$ on at most $5$ vertices, and for every largest directed rooted tree $\dT$ of $\dH$, the complement of $\dT$ in $\dH$ is a collection of rooted paths and  stars\footnote{This does not hold for subgraphs $H$ that are stars, but stars can be dealt with differently.}. Therefore, all potential completions  of a copy of $\dT$ to  $\dH$ in $\dG$ can be
computed and hashed in time $O(m\cdot poly (\alpha))$. See figure below for an illustration of a possible $\dH$ and its DRTS $\dT$ (adapted from~\cite{BPS20}).
Hence, whenever a copy of $\dT$ is discovered in $\dG$, it can be verified in $O(1/\delta)$ rounds
whether this copy can be extended to $\dH$. Since all copies of $\dT$ can be enumerated in $O(m \alpha^3)$ time,
the overall algorithm takes $O(m \alpha^3)$ time.

\begin{figure}[h!]\label{fig:DRST}
\centering
    \includegraphics[width=0.45\textwidth]{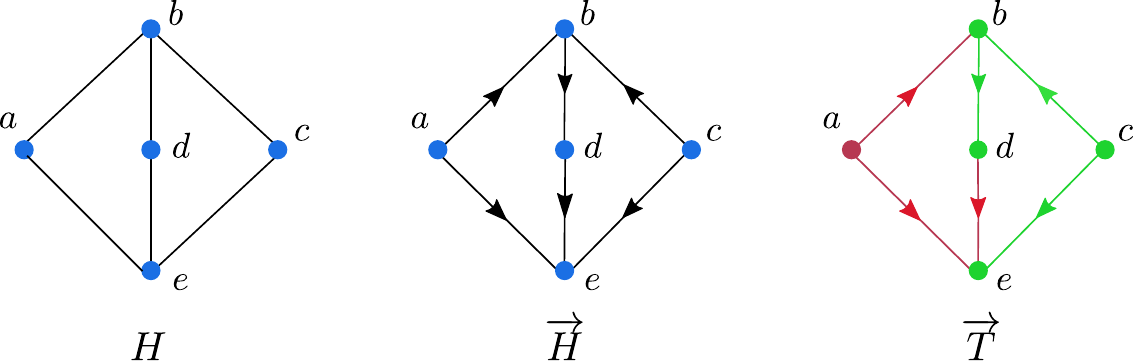}
\caption{
	From left to right:
	A subgraph H; a possible directed copy of $H$;
	the DRTS in green, and its complement with respect to $H$ in red.
    Based on a figure from BPS~\cite{BPS20}.
}

\end{figure}

\subsection{Implementation in the MPC model}\label{sec:sub-impl}

\begin{notation}[Outgoing neighbors and out-degree]
Let $\dG=(V,\dE)$ be a directed graph. For a vertex $v\in V$, We denote by $N^+(v)$ its set of \emph{outgoing neighbors}, and by $\dpp(v)=|N^+(v)|$ its \emph{outgoing degree} or \emph{out-degree}.
\end{notation}

\begin{definition}[Degeneracy and degeneracy ordering.]\label{def:degen}
A  \emph{degeneracy ordering} of a graph $G$, is an ordering obtained by repeatedly removing
a minimum degree vertex and all the edges incident to this vertex. A vertex $u$ precedes a vertex $v$ in this ordering, $u\prec v$, if $u$ was removed before $v$.
The \emph{degeneracy} of a graph $G$ is then the maximum
outgoing degree over all vertices in a degeneracy ordering of $G$.
\end{definition}

\begin{theorem}[Thm 2 in ~\cite{GLM19}.]\label{thm:GLM}
 Given a graph $G$ with arboricity $\alpha$, it outputs, with high probability, an orientation of $G$, $\dG$,  where each vertex in $\dG$ has out-degree at most $O(\alpha)$.
 The algorithm performs $O(\sqrt{\log n}\cdot \log \log n)$ rounds, uses $\widetilde{O}(n^\delta)$ space per machine,  for an arbitrary constant $\delta\in(0,1)$,  and the total memory is $O(\max\{m, n^{1+\delta}\})$.
\end{theorem}

The following is a key lemma.

\begin{lemma}\label{lem:DRTS}
	Let $\dG$ be a directed graph over $m$ edges such that each vertex has out-degree at most $\alpha$. Let $\dT$ be a directed rooted tree  of size  $t\geq 2$.
	We can list all copies of $\dT$ in $G$ in $O(1/\delta)$ rounds, $O(n^{2\delta})$ space per machine, and  $O(m\cdot \alpha^{t-2})$ total memory.
\end{lemma}
\begin{proof}
	Let $a_1, \ldots, a_{t}$ denote the vertices of $\dT$, where $a_1$ is the root, and $a_i$ is the $i^{\textrm{th}}$ vertex with respect to the BFS ordering of $\dT$. Let $\dT_i$ denote $\dT[\{a_0, \ldots, a_i\}]$.

	We prove the claim by induction on $t$.
	For $t=2$, all edges in  $G$ are copies of $\dT$, so the claim holds trivially.

	Assume that the claim holds for $i$, and we now prove it for $i+1$.
	By the assumption, in $O(1/\delta)$ rounds and $O(m\alpha^{i-2})$ total memory, all copies of $\dT_i$ can be listed.
	We will show that we can use these copies to find all copies of $\dT_{i+1}$ in $O(1/\delta)$ rounds and $O(m\alpha^{i-1})$ memory.
	Recall that we have machines with $O(n^{2\delta})$ memory.
	We will divide the copies among the machines, so that each machine only holds $O(n^{\delta})$ copies.
	Let $M$ be some machine containing copies $\tau_1, \ldots , \tau_{n^{\delta}}$ of $\dT_i$. It will be easier to think of $M$ as a collection of $n^{\delta}$ constant memory parts, each holding a single copy of $\dT_i$.
	Consider a specific copy $\tau$ of $\dT_i$ and let $P_{\tau}$ denote the part storing that copy.
	Let $a_p$ denote the vertex in $\dT$ that is the parent of $a_{i+1}$, and let $u$ denote the vertex in $\tau$ that is mapped to $a_p$.
	We would like to create all tuples $(\tau, w)$, where $w\in N^+(u)\setminus \tau$ and $w$ can be mapped to $a_{i+1}$.
	In order to achieve this we duplicate $P_{\tau}$ for $\alpha$ times, to get copies $P_{\tau, 1},\ldots, P_{\tau,\alpha}$.
	Each part $P_{(\tau,i)}$ then asks $u$ for its $i\thh$ neighbor $w$, and  then checks if $\tau$ can be extended to $\dT_{t_1+1}$ using $w$.
	If $(\tau,w)$ is a copy of $\dT_{i+1}$, then the part creates the tuple $(\tau,w)$.
All the the duplications above can be done in parallel to all copies of $\dT$ residing on a single machine, so that in total each machine is duplicated $\alpha$ time.
Since each machine has $O(n^\delta)$ information, and $O(n^{2\delta})$ space, by \cref{lem:duplicate}, this process takes $O(\log _{n^{\delta}}\alpha)=O(1/\delta)$ rounds. Furthermore, as each machine is duplicated $\alpha$ times, the amount of total memory increases by a factor of $\alpha$.

	Hence, at the end of the process, all copies of $\dT$ are generated, the round complexity is $O(1/\delta)$, and the total memory is $O(m\alpha^{t-2})$.
\end{proof}

For a directed graph $\dG$, we consider the following lists of key-value pairs, as described in Lemma 15 in~\cite{BPS20}.
\begin{itemize}\label{def:HM}
\item $\HM_1\;: \;((u,v), 1)$ for all $(u,v)\in E(\dG).$
\item $\HM_2\;:\; (S,\ell)\; \forall S\subseteq V(\dG)$ such that $1\leq |S| \leq 4$ and $\ell$ is the number of vertices $u$ such that $S\subseteq N^+(u)$.
\item  $\HM_3 \;:\;\big((S_1, S_2, \ell)\big)\; \forall S_1, S_2\subseteq V(\dG)$, where $1\leq |S_1\cup S_2|\leq 3$, and $\ell$ is the number of edges $e=(u,v) \in E(\dG)$ such that $S_1\subseteq N^+(u)$ and $S_2\subseteq N^+(v)$.
\end{itemize}

\begin{lemma}\label{lem:HM}
	Let $\dG$ be a directed graph with $m$ edges, such that for every $v\in V(\dG)$, $d^+(v)\leq \alpha$.
	The lists $\HM_1, \HM_2$ and $\HM_3$ can be computed in $O(1/\delta)$ rounds and  $O(m\alpha^3)$ total memory.
\end{lemma}

\begin{proof}
	In order to create $\HM_1$, each vertex $u$ simply adds for each $v \in N^+(u)$ the pair $((u,v),1)$ to the list. Clearly this can be done in $O(1)$ rounds, and $O(m)$ total memory.

	We now consider $\HM_2$. Let $s=|S|$ denote the size of the requested set. Fix $s$, and let $\dT$ be a DRT which consists of a root and $s$ outgoing neighbors.
	By \cref{lem:DRTS}, we can generate all copies of $\dT$ in $O(1/\delta)$ rounds,
	and $O(m\cdot \alpha^{s-2})=O(m\cdot \alpha^{2})$ total memory.
	From each copy $(v, u_1, \ldots, u_{s})$ of $\dT$, we create a tuple $(\{u_1, \ldots, u_{s}\},1)$ and add it to a temporary list $\HM_2'$.
	Finally, we use \cref{thm:mpc-interval-tree} to sort this list and aggregate the counts of each set $S=\{u_1, \ldots, u_{s}\}$, so that for every  $S$ we create the tuple $(S,\ell)$ and add it to $\HM_2$, where $\ell$ is the number of occurrences of the tuple $(S,1)$ in $\HM'_2$.
	By \cref{thm:mpc-interval-tree}, this process takes $O(\log_{n^{\delta}}m\cdot \alpha^{2})= O(1/\delta)$ rounds.

	$\HM_3$ is constructed similarly. Fix some $s_1$ and $s_2$ such that $1\leq s_1+s_2\leq 3$, and consider the corresponding DRT $\dT$. That is, $\dT$ is a DRT with a vertex $u$ with $s_1$ outgoing neighbors, where one of the neighbors has $s_2$ additional outgoing neighbors. This is a DRT over $|S_1|+|S_2|+1 \leq 4$ vertices, so by \cref{lem:DRTS}, we can generate all copies in $O(1/\delta)$ rounds, and $O(m\cdot \alpha^{2})$ total memory.
	From the list of all copies we can generate $\HM_3$, similarly to as described for $\HM_2$, in $O(1/\delta)$ rounds.
\end{proof}

\begin{theorem}\label{thm:five-subgraphs}
	Let $G=(V,E)$ be a graph with $n=|V|$ and $m=|E|$.
There is an algorithm for counting the  number of occurrences of any given  subgraph $H$ over $k\leq 5$ vertices  in $G$ with high probability, with  round complexity $O(\sqrt{\log n}+1/\delta)$, $O(n^{2\delta})$  memory per machine, and $O(m\alpha^3)$ total memory.
\end{theorem}

\begin{proof}
	If $H$ is a $k$-star, then the number of occurrence of $H$ in $G$ is simply $\sum_{v \in V}\binom{d(v)}{k}$ where $\binom{d(v)}{k}=0$ for $k> d(v)$, which can
be computed in $O(1)$ rounds.
Hence, we assume that $H$ is not a star.

	The first step in the algorithm of BPS 	is to direct the graph $G$ according to the degeneracy ordering (see \cref{def:degen}).
	We achieve this using the algorithm of~\cite{GLM19} described in \cref{thm:GLM}. Note that the algorithm of~\cite{GLM19} returns an approximate degeneracy ordering, but as the degeneracy of a graph is at most twice the arboricity, it holds that each vertex has out-degree $O(\alpha)$.

	Given the ordering of $\dG$, the algorithm continues by considering all orientations $\dH$ of $H$ (up to isomorphisms). For each $\dH$ it computes the maximal rooted directed tree, DRT, of $\dH$, denoted $\dT$. As $H$ is of constant size, this
	can be computed in $O(1)$ rounds on a single machine.

	The next step is to find all copies of $\dT$ in $\dG$.
	By \cref{lem:DRTS},  this can be implemented in $O(1/\delta)$ rounds, $O(n^{2\delta})$ space per machine, and $O(m\alpha^2)$ total memory.

	Now, for each copy of $\dT$ in $\dG$ it needs to be verified if the copy can be completed to a copy of $\dH$ in $\dG$.
	 By Lemma 16 in~\cite{BPS20}, this can be computed in if given query access to $\HM_1, \HM_2$ and $\HM_3$, as defined in \cref{def:HM}.
	 That is, it can be determined if a copy $\tau$ of $\dT$ using $O(|H|^2)=O(1)$  queries to the lists $\HM_1$, $\HM_2$ and $\HM_3$.
	 By \cref{lem:HM}, these lists can be generated in $O(1/\delta)$ rounds, and $O(m\alpha^2)$ total memory.
	 For $i\in [1..3]$, let $Q_i$ denote the set of all queries to list $\HM_i$.
	  By~\cite{GSZ11},  all  queries $Q_i$ to $\HM_i$ can be answered in time $O(1/\delta)$.

	 Finally, by Lemma 16 in~\cite{BPS20}, each $v$ can use the answers to its queries to compute the number of copies of $\dH$ it can be extended to. Therefore, by summing over all vertices and over all possible orientations of $H$, and taking into account isomorphisms, we can compute the number of occurrences of $H$ in $\dG$.
	 The total round complexity is dominated by computing the approximate arboricity orientation of $G$ and the sorting operations. Therefore the round complexity is $O(\sqrt{\log n}\log\log n+1/\delta)$. The space per machine is $O(n^{2\delta})$, and the total memory over all machines is $O(m\alpha^3)$.
\end{proof}

%% file: 590-experiments.tex
\pgfplotsset{
    bar group size/.style 2 args={
        /pgf/bar shift={%
                -0.5*(#2*\pgfplotbarwidth + (#2-1)*\pgfkeysvalueof{/pgfplots/bar group skip})  +
                (.5+#1)*\pgfplotbarwidth + #1*\pgfkeysvalueof{/pgfplots/bar group skip}},%
    },
    bar group skip/.initial=2pt,
    plot0/.style={blue,fill=blue!30!white,mark=none},%
    plot1/.style={red,fill=red!30!white,mark=none},%
    plot2/.style={brown!60!black,fill=brown!30!white,mark=none},%
    plot3/.style={green!60!black,fill=green!30!white,mark=none},%
    plot4/.style={purple!60!black,fill=purple!30!white,mark=none},%
    plot5/.style={orange!60!black,fill=orange!30!white,mark=none},%
}

\section{Experiments}

We performed experiments using our algorithms
given in \cref{sec:approx-triangle-counting,sec:exact_triangle_counting}.
Our code~\cite{experiments-code} simulates the algorithms
described in these sections as well as the MPC procedures
we use as subroutines.
In the  implementation of the algorithm of \cref{sec:approx-triangle-counting},
our algorithm achieves a better approximation \emph{on all tested graphs} than
the best-known previous algorithm.
In the implementation of the exact algorithm of
\cref{sec:exact_triangle_counting},
our algorithm achieves fewer number of MPC rounds than the baseline algorithm.
Given that the focus of this paper is on our theoretical contributions, we
include these experiments only as a proof-of-concept below.
All real-world graphs on which we performed our experiments can be
found in the Stanford Large Network Dataset Collection (SNAP)~\cite{snapnets}.
We leave as interesting future directions
implementing and testing our algorithms in massively parallel
software framworks, such as Apache Hadoop and others,
on much larger graphs.

%% file: 700-experiments-appendix.tex
\pgfplotsset{
    bar group size/.style 2 args={
        /pgf/bar shift={%
                -0.5*(#2*\pgfplotbarwidth + (#2-1)*\pgfkeysvalueof{/pgfplots/bar group skip})  +
                (.5+#1)*\pgfplotbarwidth + #1*\pgfkeysvalueof{/pgfplots/bar group skip}},%
    },
    bar group skip/.initial=2pt,
    plot0/.style={blue,fill=blue!30!white,mark=none},%
    plot1/.style={red,fill=red!30!white,mark=none},%
    plot2/.style={brown!60!black,fill=brown!30!white,mark=none},%
    plot3/.style={green!60!black,fill=green!30!white,mark=none},%
    plot4/.style={purple!60!black,fill=purple!30!white,mark=none},%
    plot5/.style={orange!60!black,fill=orange!30!white,mark=none},%
}

We performed experiments using our algorithms
given in \cref{sec:approx-triangle-counting,sec:exact_triangle_counting}.
Our code~\cite{experiments-code} simulates the algorithms
described in these sections as well as the MPC procedures
we use as subroutines.
In the  implementation of the algorithm of \cref{sec:approx-triangle-counting}
we output the approximation factor we achieve using our algorithm versus
the amount of space per machine.
In the implementation of the exact algorithm of \cref{sec:exact_triangle_counting}
we output the number of MPC rounds necessary to execute the algorithm
versus the arboricity bound we pass into it.
Given that the focus of this paper is on our theoretical contributions, we
include these experiments only as a proof-of-concept.
We leave as interesting future directions
implementing and testing our algorithms in massively parallel
software framworks, such as Apache Hadoop and others,
on much larger graphs.

All real-world graphs on which we performed our experiments can be
found in the Stanford Large Network Dataset Collection (SNAP)~\cite{snapnets}.

We test our algorithms against datasets described in \cref{table:datasets}.

\begin{table}[htb!]\label{table:files}
\centering
\resizebox{0.7\textwidth}{!}{%
\begin{tabular}{cccc}
\toprule
File        & Number of Vertices ($n$)          & Number of Edges ($m$)       & Number of Triangles ($T$) \\
\midrule
\textbf{email-Eu-core} & 1005 & 25571 & 105461\\
\midrule
\textbf{ego-Facebook} & 4039 & 88234 & 1612010\\
\midrule
\textbf{feather-lastfm-social} & 7624 & 27806 & 40433\\
\midrule
\textbf{ca-GrQc} & 5242 & 14496 & 48260 \\
\midrule
\textbf{musae-twitch (DE)} & 9498 & 153138 & 603088\\
\midrule
\textbf{ca-HepTh} & 9877 & 25998 & 28339 \\
\midrule
\textbf{oregon1\_010519} & 11051 & 22724 & 17677\\
\midrule
\textbf{ca-HepPh} & 12008 & 118521 & 3358499\\
\midrule
\textbf{email-Enron} & 36692 & 183831 & 727044\\
\bottomrule
\end{tabular}}
\caption{All datasets can be found in the Stanford Large
Network Dataset (SNAP) Collection~\cite{snapnets}. This table
shows the number of vertices, edges, and exact number of triangles
in each of these graphs.}\label{table:datasets}
\end{table}

\myparagraph{Results for \cref{sec:exact_triangle_counting}} The first set of experiments were performed using our new exact
triangle count algorithm provided in \cref{sec:exact_triangle_counting}.
The results  are shown in~\cref{fig:arb-exp}.
Our experiments were performed on five datasets: \textsf{oregon1\_010519, email-Enron,
ca-HepTh, ca-GrQc, email-Eu-core}. We compare against the baseline
algorithm (labeled ``-base'' in the figures) of removing
(and counting) the vertices with degree at most the degeneracy
of each graph during each round. We measure the number of rounds
our algorithm takes against the amount of space per machine (indicated
by the different colors of the bars) and our \emph{initial} setting of our
degree bound. Recall that since the degree
bound of our algorithm for removal of vertices grows doubly exponentially, we
can set our initial degree bound to be smaller than the degeneracy of the graph. Note that for the baseline algorithm, we cannot do this
since the degree bound remains the same (and hence, a smaller
    degree bound than the degeneracy will cause the algorithm
    to terminate at some point with vertices still left
in the graph with degree greater than the bound). Our
initial degree bound settings are shown on the x-axes
while the y-axes shows the number of rounds. The machine
space bounds are in terms of the \emph{number of nodes}
of the graph that can fit in each machine. These numbers are
shown in the legend indicating the colors of the bars.
Due to its usefulness in the baseline algorithm, we consider the degeneracy
of the graphs instead of the arboricity. But as we noted previously,
the degeneracy of the graph is at most within a constant factor of $2$
of the arboricity of the graph.

We see in~\cref{fig:arb-exp} that our algorithms result in less
(or the same number of) rounds than the baseline algorithm for all cases given the same
space and degree bound except the 2225 case for the \textsf{email-Enron}
dataset. This confirms our theoretical analysis of the asymptotic
number of rounds of our algorithm, taking $O_{\delta}(\log \log n)$ compared
to the $O_{\delta}(\log n)$ rounds we expect the baseline algorithm
to take. The anomaly with the single 2225 case might be due to the larger
constant factor (derived from MPC sort and find duplicate)
of having to sort more items per round in our case
compared to the baseline case. We also see that due to the doubly exponential
nature of our algorithm, we are able to also count triangles using
much lower initial degree bounds than the degeneracy of the graph.
Even for such degree bounds, the number of rounds necessary is
still often (much) less than the number of rounds necessary
for the baseline algorithm.
This provides the advantage of being able to use our algorithm
for real-world graphs without first needing to determine their
degeneracy value.

\begin{figure*}[htp!]
\centering
\begin{subfigure}{0.41\textwidth}
\pgfplotstableread[row sep=\\,col sep=&]{
    interval & 16 & 105 & 4355 & 16-arb & 105-arb & 4355-arb \\
    5     & 23  & 17  & 7 & & &\\
    10     & 19 & 13  & 7 & & & \\
    17     & 16 & 12 & 6  & 57 & 31 & 13 \\
    }\mydata
\begin{tikzpicture}
    \begin{axis}[
            ybar,
            bar width=6,
            width=\textwidth,
            height=0.8\textwidth,
            legend style={at={(0.5,1.2)},
                anchor=north,legend columns=-1},
                symbolic x coords={5, 10, 17},
            xtick=data,
            nodes near coords,
            nodes near coords align={vertical},
            ymin=0,ymax=60,
            ylabel={Rounds},
            xlabel={Degree Bounds},
            enlarge y limits={0.25,upper},
            enlarge x limits=0.25,
            ylabel shift=2pt,
            legend image code/.code={%
            \draw[#1] (0cm,-0.1cm) rectangle (0.1cm,0.1cm);}
        ]
        \addplot[plot0] table[x=interval,y=16]{\mydata};
        \addplot[plot1] table[x=interval,y=105]{\mydata};
        \addplot[plot2] table[x=interval,y=4355]{\mydata};
        \addplot[plot3] table[x=interval,y=16-arb]{\mydata};
        \addplot[plot4] table[x=interval,y=105-arb]{\mydata};
        \addplot[plot5] table[x=interval,y=4355-arb]{\mydata};
        \legend{16, 105, 4355, 16-base, 105-base, 4355-base}
    \end{axis}
\end{tikzpicture}
\caption{oregon1\_010519. Degeneracy is $17$.}
\end{subfigure}
    \begin{subfigure}{0.41\textwidth}
\pgfplotstableread[row sep=\\,col sep=&]{
    interval & 13 & 72 & 2225 & 72-arb & 2225-arb \\
    5     & 23  & 15  & 2 & & \\
    10     & 18 & 10  & 1 & & \\
    43     & 9 & 7 & 5  & 15 & 2\\
    }\mydata

\begin{tikzpicture}
    \begin{axis}[
            ybar,
            bar width=6,
            width=\textwidth,
            height=0.8\textwidth,
            legend style={at={(0.5,1.2)},
                anchor=north,legend columns=-1},
            symbolic x coords={5,10,43},
            xtick=data,
            nodes near coords,
            nodes near coords align={vertical},
            ymin=0,ymax=25,
            ylabel={Rounds},
            xlabel={Degree Bounds},
            enlarge y limits={0.25,upper},
            enlarge x limits=0.25,
            ylabel shift=1.5pt,
            legend image code/.code={%
            \draw[#1] (0cm,-0.1cm) rectangle (0.1cm,0.1cm);}
            ylabel style={yshift=0.4cm},
        ]
        \addplot[plot0] table[x=interval,y=13]{\mydata};
        \addplot[plot1] table[x=interval,y=72]{\mydata};
        \addplot[plot2] table[x=interval,y=2225]{\mydata};
        \addplot[plot3] table[x=interval,y=72-arb]{\mydata};
        \addplot[plot4] table[x=interval,y=2225-arb]{\mydata};
        \legend{13, 72, 2225, 72-base, 2225-base}
    \end{axis}
\end{tikzpicture}
\caption{email-Enron. Degeneracy is $43$.}
\end{subfigure}
\begin{subfigure}{0.4\textwidth}
\pgfplotstableread[row sep=\\,col sep=&]{
    interval & 15 & 99 & 3936 & 99-arb & 3936-arb \\
    5     & 21  & 15  & 2 & & \\
    10     & 14 & 6  & 1 & & \\
    31     & 9 & 5 & 1  & 10 & 1\\
    }\mydata

\begin{tikzpicture}
    \begin{axis}[
            ybar,
            bar width=6,
            width=\textwidth,
            height=0.8\textwidth,
            legend style={at={(0.5,1.2)},
                anchor=north,legend columns=-1},
            symbolic x coords={5,10,31},
            xtick=data,
            nodes near coords,
            nodes near coords align={vertical},
            ymin=0,ymax=25,
            ylabel={Rounds},
            xlabel={Degree Bounds},
            enlarge y limits={0.25,upper},
            enlarge x limits=0.25,
            ylabel shift=1.5pt,
            legend image code/.code={%
            \draw[#1] (0cm,-0.1cm) rectangle (0.1cm,0.1cm);}
        ]
        \addplot[plot0] table[x=interval,y=15]{\mydata};
        \addplot[plot1] table[x=interval,y=99]{\mydata};
        \addplot[plot2] table[x=interval,y=3936]{\mydata};
        \addplot[plot3] table[x=interval,y=99-arb]{\mydata};
        \addplot[plot4] table[x=interval,y=3936-arb]{\mydata};
        \legend{15, 99, 3936, 99-base, 3936-base}
    \end{axis}
\end{tikzpicture}
\caption{ca-HepTh. Degeneracy is $31$.}
\end{subfigure}
\begin{subfigure}{0.4\textwidth}
\pgfplotstableread[row sep=\\,col sep=&]{
    interval & 191 & 12826 & 191-arb & 12826-arb \\
    10     & 20  & 7  &  & \\
    20     & 19 & 11  &  & \\
    43     & 12 & 6 & 145  & 28\\
    }\mydata
\begin{tikzpicture}
    \begin{axis}[
            ybar,
            bar width=6,
            width=\textwidth,
            height=0.8\textwidth,
            legend style={at={(0.5,1.2)},
                anchor=north,legend columns=-1},
                symbolic x coords={10, 20, 43},
            xtick=data,
            nodes near coords,
            nodes near coords align={vertical},
            ymin=0,ymax=150,
            ylabel={Rounds},
            xlabel={Degree Bounds},
            enlarge y limits={0.25,upper},
            enlarge x limits=0.25,
            ylabel shift=1.5pt,
            legend image code/.code={%
            \draw[#1] (0cm,-0.1cm) rectangle (0.1cm,0.1cm);}
        ]
        \addplot[plot1] table[x=interval,y=191]{\mydata};
        \addplot[plot2] table[x=interval,y=12826]{\mydata};
        \addplot[plot3] table[x=interval,y=191-arb]{\mydata};
        \addplot[plot4] table[x=interval,y=12826-arb]{\mydata};
        \legend{191, 12826, 191-base, 12826-base}
    \end{axis}
\end{tikzpicture}
\caption{ca-GrQc. Degeneracy is $43$.}
\end{subfigure}
\begin{subfigure}{0.4\textwidth}
\pgfplotstableread[row sep=\\,col sep=&]{
    interval & 7 & 31 & 494 & 31-arb & 494-arb \\
    2     & 60  & 32  & 17 & nan & nan\\
    10     & 39 & 21  & 15 & nan & nan \\
    34     & -1 & 14 & 6  & 89 & 43\\
    }\mydata
\begin{tikzpicture}
    \begin{axis}[
            ybar,
            bar width=6,
            width=\textwidth,
            height=0.8\textwidth,
            legend style={at={(0.5,1.2)},
                anchor=north,legend columns=-1},
                symbolic x coords={2, 10,34},
            xtick=data,
            nodes near coords,
            nodes near coords align={vertical},
            ymin=0,ymax=90,
            ylabel={Rounds},
            xlabel={Degree Bounds},
            enlarge y limits={0.25,upper},
            enlarge x limits=0.25,
            unbounded coords = jump,
            ylabel shift=2pt,
            legend image code/.code={%
            \draw[#1] (0cm,-0.1cm) rectangle (0.1cm,0.1cm);}
        ]
        \addplot[plot0] table[x=interval,y=7, area legend]{\mydata};
        \addplot[plot1] table[x=interval,y=31, area legend]{\mydata};
        \addplot[plot2] table[x=interval,y=494, area legend]{\mydata};
        \addplot[plot3] table[x=interval,y=31-arb, area legend]{\mydata};
        \addplot[plot4] table[x=interval,y=494-arb, area legend]{\mydata};
        \legend{7, 31, 494, 31-base, 494-base}
    \end{axis}
\end{tikzpicture}
\caption{email-Eu-core. Degeneracy is $34$.}
\end{subfigure}
\caption{This set of graphs shows the results of our experiments
    using our exact counting algorithm described in~\cref{sec:exact_triangle_counting}.
    We test on five datasets labeled under each plot. In each of these graphs,
    we compare against the number of rounds required by the MPC algorithm that
    removes, in each round, only vertices with degree at most the degeneracy
    of the graph $\alpha$.
    Each color represents a different space per machine, which is represented in terms of the number of nodes that
    can fit in each machine. The colors (green, red, yellow)
    labeled with ``-base'' represent
our baseline algorithm results.}\label{fig:arb-exp}
\end{figure*}

\begin{table}[htb!]\label{table:files}
\centering
\resizebox{0.7\textwidth}{!}{%
\begin{tabular}{cccc}
\toprule
  File        & $\delta$      & Partition Approximation   & Our Approximation \\
\midrule
\textbf{ego-Facebook} & 0.5 & 0.62 & 1.31\\
\midrule
\textbf{feather-lastfm-social} & 0.5 & 5.41 & 1.08\\
\midrule
\textbf{ca-GrQc} & 0.5 & 4.53 & 1.64 \\
\midrule
\textbf{ca-HepPh} & 0.5 & 0.66 & 1.22\\
\midrule
\textbf{ego-Facebook} & 0.75 & 0.75 & 0.97\\
\midrule
\textbf{ca-GrQc} & 0.75 & 5.82 & 0.82 \\
\midrule
\textbf{ca-HepPh} & 0.75 & 5.90 & 0.86\\
\midrule
\textbf{musae-twitch (DE)} & 0.75 & 0.74 & 0.95\\
\midrule
\textbf{oregon1\_010519} & 0.75 & 0.60 & 0.71\\
\bottomrule
\end{tabular}}
\caption{The approximation factors obtained
    when running our algorithm given in~\cref{sec:approx-triangle-counting}
    against our implementation of the partition algorithm given in
    Algorithm 1 and Algorithm 2 of~\cite{pagh2012colorful}. We perform
the algorithms on machines of size $2m^{\delta} \cdot \log{n}$. The
approximation factor is calculated by $C/T$ where $C$ is the triangle
count returned by either algorithm and $T$ is the actual count of
the triangles in the graph.}\label{table:approx}
\end{table}

\paragraph{Results for \cref{sec:approx-triangle-counting}.} The results of experiments for our approximation algorithm
described in \cref{sec:approx-triangle-counting} are given in \cref{table:approx}.
We further compare our approximations against our implementation
of the partition algorithm given in Algorithms~1 and~2 of~\cite{pagh2012colorful}.
In the implementation of our algorithm, due to the (sequential) time constraints
of simulating our algorithms, we do not use a $k$-wise independent
hash function, as such functions require too much time to compute.
Hence, for our experiments, we use standard pseudorandom functions
given in programming packages (specifically \textsf{numpy} in
\textsf{Python3}).
For each of the experiments the space per machine is $2m^{\delta} \cdot \log{n}$,
where $\delta$ is specified in~\cref{table:approx}.
The same space per machine is used for both algorithms.
The total space used for both
algorithms is $2m \cdot \log{m}$.
For the implementation of our algorithm, we set the
probability of sampling to be $\frac{1}{5} \cdot \sqrt{\frac{S}{M\cdot k}}$
where we set $k = \max(2, \floor{6.5 \cdot \log{n}})$.
We chose to test these algorithms
on these specific $\delta$ values because $\delta = 0.5, 0.7$
represent $\tilde{O}(n)$ and $\tilde{o}(m)$, respectively.
Because the theoretical guarantees of our algorithm
relies on some specific contraints on $T$ and
$S$, we wanted to see how our algorithm performs on real-world
networks. We use the median-of-means trick for the concentration
for both algorithms.

As \cref{table:approx} shows, compared to the partition algorithm, our algorithm obtains a better
approximation ratio for all datasets and for all machine spaces.
This follows from our theoretical analysis
as we ensure $(1+\eps)$-approximations on the number of triangles
in each graph using $\tilde{O}(n)$ space and with a quadratically
smaller constraint on the number of actual triangles in the graph
than all other previous work. Thus, we show that practically, on real-world graphs,
our algorithm obtains better approximations on the number of triangles
given smaller space per machine.

%% file: 1200-appendix-preliminaries.tex
\section{Preliminaries}
\label{sec:appendix-prelims}

\subsection{Proof of \cref{thm:mpc-interval-tree}}
Using the construction of the interval tree defined in~\cite{GSZ11} that has branching
factor $d = \mM/2$ we perform the following to count the number of times each element repeats in
our sorted list of $N$ elements. To initialize the tree, each leaf of the tree contains
exactly one of the elements in the sorted list of elements where leaf $v_{i}$ contains
element $x_i$ of the list. Let the height of the
tree be $L$, the leaves of the tree be at level $L - 1$ and
the root be at level $0$. Then, the rest of the algorithm proceeds in two phases:
\begin{enumerate}
    \item \textbf{Bottom-up phase:} For each level $\ell = L-1$ up to $0$:
    \begin{enumerate}
        \item For each node $v$ on level $\ell$:
        \begin{enumerate}
            \item If $v$ is a leaf, it sends its value $x_i$ to its parent $p(v)$.
            \item If $v$ is a vertex in level $L - 2$, let $(x_i, x_{i+1}, \dots, x_{i +j})$ where $j < d$ be values obtained from its leaf children
            from left to right. Let $c(x_i)$ be the count of element $x_i$ among the values obtained from the children of $v$. The counts are computed
            locally on the machine storing $v$. Then, $v$ sends $x_i$, $c(x_i)$, $x_{i+j}$, $c(x_{i+j})$ to its parent $p(v)$.
            \item If $v$ is a non-leaf node on level $\ell < L-2$, let $x_a, c(a), x_b, c(b), \dots$ be the values
            of elements obtained from its children and their counts. $v$ updates the counts of all elements received. For example,
            if $x_a = x_b$, $v$ updates $c(a)$ and $c(b)$ to be $c(a) + c(b)$. Let $x_{left}$ be the first element received
            from $v$'s leftmost leaf and $x_{right}$ be the second element received from $v$'s rightmost leaf. Then, send these elements and their updated counts,
            $x_{left}$, $c(x_{left})$, $x_{right}$, and $c(x_{right})$, to its parent $p(v)$.
        \end{enumerate}
    \end{enumerate}
    \item \textbf{Top-down phase:} For each level $\ell = 0$ down to $\ell = L-1$:
    \begin{enumerate}
        \item For each node $v$ at level $\ell$:
        \begin{enumerate}
            \item If $v$ is the root, then it computed and stored in its memory new repeating counts for the values it received from its children: $x_a, c(x_a), x_b, c(x_b), \dots$.
            It sends the new counts and values to its respective child that sent it the value originally (e.g.\ $x_{left}, c(x_{left})$ to $v_{left}$). Intuitively, this updates
            the child's count of values with values that are not in its subtree.
            \item If $v$ is not the root and is a non-leaf node, it receives the values from its parents for its leftmost and rightmost child counts.
            Given the set of values it stored from its children
            it updates the counts with counts of values received from its parents. This allows for the counts to reflect values
            not in its subtree. Then, it sends the updated counts to its children.
            \item If $v$ is a leaf, it receives values $x_i$, $c(x_i)$ from its parent.
            $c(x_i)$ is then the number of times $x_i$ occurs in the
            sorted list.
        \end{enumerate}
    \end{enumerate}
\end{enumerate}
The above procedure uses $O(d)$ space per processor and $O(L)$ rounds of communication.
Since $L = O(\log_d(N))$ and $d = \mM/2$, the number of rounds of communication that is
necessary is $O\left(\log_{\mM} N\right)$.

\subsection{Proof of \cref{lem:queries}}

We first create the following tuples in parallel
to represent tuples in $Q$ and $C$, respectively.
For each tuple $q \in Q$, we create the tuple $(q, 1)$.
For each tuple $c \in C$, we create the tuple $(c, 0)$.
Let $F$ denote the set of tuples $(c, 0)$ and $(q, 1)$. First,
we sort the tuples in $F$ lexicographically (where $0$ comes before $1$)~\cite{GSZ11}.
Then, we use the predecessor primitive given in (e.g.\ \cite{GSZ11,ASSWZ18}, Appendix A of~\cite{BDELM19})
to determine the queries $q\in Q$ that are in $C$.
Given the sorted $F$, we use the predecessor algorithm of~\cite{BDELM19}
to determine for each $(q, 1)$ tuple, the first tuple that appears before it that
has value $0$. Suppose this tuple is $(c, 0)$. Then, if $q = c$, then
the queried tuple $q$ is in $C$. For all tuples $q \in Q$, we can
then return in parallel whether $q \in C$ also. Both the
sorting and the predecessor queries take $O(|Q \cup C|)$ total space
and $\hideo(1)$ rounds.

\subsection{Proof of \cref{lem:duplicate}}
	Let $M$ be some machine with $n^{\delta}$  information and $O(n^{2\delta})$ space.  We create the $x$ duplicates by repeatedly   duplicating each machine $M^i_j$ to $n^{\delta}$ machines $M^{i+1}_{n^{\delta}\cdot j}, \ldots, M^{i+1}_{n^{\delta}\cdot j+n^{\delta}-1}$, starting with $M^0_0=M$. Therefore, after $\ell=\log _{n^{\delta}}x$ rounds this process terminates, and the required duplicates is the set of machines $M^{\ell}_1$ to $M^{\ell}_x$.